\documentclass[a4paper,11pt]{article}
\pdfoutput=1
\usepackage{jcappub}
\usepackage[T1]{fontenc} 
\usepackage{hyperref}
\usepackage{graphics}
\usepackage{amssymb, amsmath}

\usepackage{xcolor}

\author[a,b,c]{Martin White}
\author[b]{Rongpu Zhou}
\author[b]{Joseph DeRose}
\author[b]{Simone Ferraro}
\author[a,b]{Shi-Fan Chen}
\author[d]{Nickolas Kokron}
\author[b]{Stephen Bailey}
\author[e]{David Brooks}
\author[f]{Juan Garc\'ia-Bellido}
\author[b]{Julien Guy}
\author[g]{Klaus Honscheid}
\author[h]{Robert Kehoe}
\author[b]{Anthony Kremin}
\author[b]{Michael Levi}
\author[b]{Nathalie Palanque-Delabrouille}
\author[b]{Claire Poppett}
\author[b]{David Schlegel}
\author[i]{Gregory Tarle}

\affiliation[a]{Department of Physics, University of California, Berkeley, CA 94720}
\affiliation[b]{Physics Division, Lawrence Berkeley National Laboratory, Berkeley, CA 94720}
\affiliation[c]{Department of Astronomy, University of California, Berkeley, CA 94720}
\affiliation[d]{Kavli Institute for Particle Astrophysics and Cosmology and Department of Physics, Stanford University, 382 Via Pueblo Mall, Stanford, CA 94305}
\affiliation[e]{Department of Physics and Astronomy, University College London, WC1E 6BT UK}
\affiliation[f]{Instituto de F\'isica Te\'orica UAM-CSIC, Universidad Aut\'onoma de Madrid , 28049 Madrid, Spain}
\affiliation[g]{Ohio State University, Columbus, OH 43210}
\affiliation[h]{Department of Physics, Southern Methodist University, Dallas, TX 75275}
\affiliation[i]{Department of Physics, University of Michigan, Ann Arbor, MI 48109}

\emailAdd{mwhite@berkeley.edu}
\emailAdd{rongpuzhou@lbl.gov}
\emailAdd{joe.derose13@gmail.com}
\emailAdd{sferraro@lbl.gov}
\emailAdd{shifan\_chen@berkeley.edu}
\emailAdd{kokron@stanford.edu}

\title{Cosmological constraints from the tomographic cross-correlation of DESI Luminous Red Galaxies and Planck CMB lensing}


\abstract{We use luminous red galaxies selected from the imaging surveys that are being used for targeting by the Dark Energy Spectroscopic Instrument (DESI) in combination with CMB lensing maps from the Planck collaboration to probe the amplitude of large-scale structure over $0.4\le  z\le 1$.  Our galaxy sample, with an angular number density of approximately $500\,\mathrm{deg}^{-2}$ over 18,000 sq.deg., is divided into 4 tomographic bins by photometric redshift and the redshift distributions are calibrated using spectroscopy from DESI.  We fit the galaxy autospectra and galaxy-convergence cross-spectra using models based on cosmological perturbation theory, restricting to large scales that are expected to be well described by such models.  Within the context of $\Lambda$CDM, combining all 4 samples and using priors on the background cosmology from supernova and baryon acoustic oscillation measurements, we find $S_8=\sigma_8(\Omega_m/0.3)^{0.5}=0.73\pm 0.03$.  This result is lower than the prediction of the $\Lambda$CDM model conditioned on the Planck data.  Our data prefer a slower growth of structure at low redshift than the model predictions, though at only modest significance. }
\arxivnumber{2111.09898}

\begin{document}
\maketitle
\flushbottom

\section{Introduction}

The growth of the large-scale structure of the Universe from small perturbations laid down at early times, as probed by fluctuations in the cosmic microwave background, to the galaxies, clusters and superclusters of the modern era is a prediction of our standard cosmological model.  Measuring the growth rate and comparing to this prediction provides a strong test of many facets of this theory, including the contents, expansion history and laws of gravity in our Universe.  One means of measuring the inhomogeneity in the modern Universe is through gravitational lensing.  In particular, gravitational lensing of CMB photons provides us with a relativistic tracer that gives an unbiased measure of the gravitational potentials associated with large-scale structure (see refs.~\cite{Lewis06,Hanson10} for reviews).  With the statistical properties and redshift of the source (primary CMB anisotropies) extremely well characterized, CMB lensing is a robust probe and one that is rapidly increasing in statistical power.

Like other lensing probes CMB lensing is a projected signal, sensitive to a wide range of redshifts, and this makes isolating a particular epoch difficult.  By cross-correlating CMB lensing with low-redshift tracers (such as galaxies) that can be confined to narrow slices in redshift we can extract the information as a function of redshift (lensing tomography). Moreover, by combining tracers with different dependences on galaxy bias, we can efficiently break the degeneracy between bias and the amplitude of fluctuations at low redshift, providing tight cosmological constraints on the low-redshift Universe as a function of epoch.

While the $\Lambda$CDM model, conditioned on CMB data, makes highly precise predictions for the amplitude of large-scale structure ($\sigma_8$) as a function of redshift in the low redshift Universe \cite{PlanckLegacy18} the observational situation as regards testing these predictions is currently mixed (see also \S\ref{sec:comparison}).  Some inferences from the abundance of clusters suggest lower values of $\sigma_8$ than the CMB-based model predicts (see \S 6.4 of ref.~\cite{PlanckLegacy18} or ref.~\cite{Bolliet18} for discussion), while other inferences from the thermal Sunyaev-Zeldovich (tSZ; \cite{SZ72}) power spectrum suggest values in good agreement with CMB \cite{Horowitz17} depending on assumptions about the relation between the observable and halo mass.  Surveys measuring cosmic shear tend to find lower values than high redshift inferences \cite{Heymans:2013fya,kids1000,Hikage:2018qbn,DESY3,Amon21,Secco21,Loureiro21} while studies of velocities of galaxies via redshift-space distortions find ``low'' \cite{Ivanov20,DAmico20,Colas:2019ret,Troster:2019ean,Chen22,Ivanov21}, ``intermediate'' \cite{Zhang21b,Kobayashi21}  and ``high'' \cite{eBOSS:2020yzd} results.  Earlier analyses of galaxies, clusters and quasars in combination with Planck lensing \cite{PlanckLens13,PlanckLens15,PlanckLens18} found either consistent \cite{PlanckLens13} or lower \cite{Pullen16,Doux18,Singh20,Hang21a,Hang21b,Kitanidis21} clustering amplitudes than expected from CMB, though with low statistical significance.  By contrast the combination of BOSS galaxies at $z\approx 0.6$ and the ACT lensing maps is consist with the CMB-preferred normalization \cite{Darwish21}.  In a recent analysis of a wide range of data sets, ref.~\cite{Garcia21} find evidence for a deficit in the growth of large scale structure in the redshift regime $0.2<z<0.6$.

Given this situation, the purpose of this paper is to measure the amplitude of large scale structure for $z\simeq 0.5-1$ by combining recently acquired samples of luminous red galaxies covering 18,000 sq.deg.\ of the extragalactic sky \cite{Dey19} with the CMB lensing map probed by Planck \cite{PlanckLens18}. 
In a companion paper \cite{PaperI} we present the samples of galaxies we use, drawn from the imaging surveys \cite{Dey19} being used for targeting by the Dark Energy Spectroscopic Instrument (DESI; \cite{DESI}).  These samples included those being targeted by DESI for follow-up spectroscopy, split by photometric redshift into different redshift ranges \cite{PaperI}.  In this paper we present the cross-correlation between those maps and the Planck 2018 CMB lensing maps \cite{PlanckLens18} and provide an initial cosmological investigation.  A particular goal is to measure the amplitude of fluctuations, $\sigma_8$, using these samples.  In the future we will be able to combine these lensing inferences (or those with higher signal-to-noise ratio lensing maps) with constraints from galaxy-galaxy lensing, baryon acoustic oscillations and redshift-space distortions measured from the same samples of galaxies and thus provide a more comprehensive test of our cosmological model.

The outline of the paper is as follows.  In \S\ref{sec:data} we review the data going into our analysis including computation of the (pseudo-)spectra, window functions and covariances.  We describe the modeling in \S\ref{sec:modeling} and tests of our pipeline in \S\ref{sec:simulations}.  Our results are presented in \S\ref{sec:results} while \S\ref{sec:conclusions} contains our conclusions and directions for future work.  We use units where $c=1$ throughout unless otherwise specified.  All distances are comoving and in $h^{-1}$Mpc units.

\section{Data}
\label{sec:data}

\subsection{Luminous Red Galaxies}

\begin{table}[]
    \centering
    \begin{tabular}{c|ccccccccc}
        Sample & $\bar{z}$ & $\Delta z$ & $\bar{n}_\theta$ & $10^6\,$SN&$s_\mu$& $z_{\rm eff}$    & $\ell_{\rm max}$& $b_E$& $\ell_{SN}$ \\ \hline
         1     &  0.47     & 0.064      &  83              & 4.02      & 0.982 & 0.47             & 250             & 1.9  & 400 \\
         2     &  0.63     & 0.075      & 149              & 2.24      & 1.033 & 0.62             & 300             & 2.1  & 530 \\
         3     &  0.79     & 0.079      & 162              & 2.07      & 0.961 & 0.78             & 350             & 2.4  & 575 \\
         4     &  0.92     & 0.097      & 149              & 2.26      & 1.013 & 0.91             & 400             & 2.5  & 425
    \end{tabular}
    \caption{Some properties of the samples we analyze.  The sample number refers to the {\tt pz\_bin} index in the catalogs \cite{PaperI}.  The redshift location and scale of each sample are provided, as well as the mean number of galaxies per square degree ($\bar{n}_\theta$) and the Poisson shot-noise level, SN.  The (log)slope of the magnitude distribution, $s_\mu$, is used for the magnification calculation (Eq.~\ref{eqn:sdef}) while the effective redshift is defined in Eq.~(\ref{eqn:zeff}).   In the last three columns we give the maximum $\ell$ used in the fits (\S\ref{sec:pk_models}), the inferred large-scale (Eulerian) bias ($b$; \S\ref{sec:results}) and the $\ell$ at which Poisson shot noise equals the cosmological power in the best-fitting auto-spectrum model. }
    \label{tab:samples}
\end{table}

\begin{figure}
    \centering
    \resizebox{0.8\columnwidth}{!}{\includegraphics{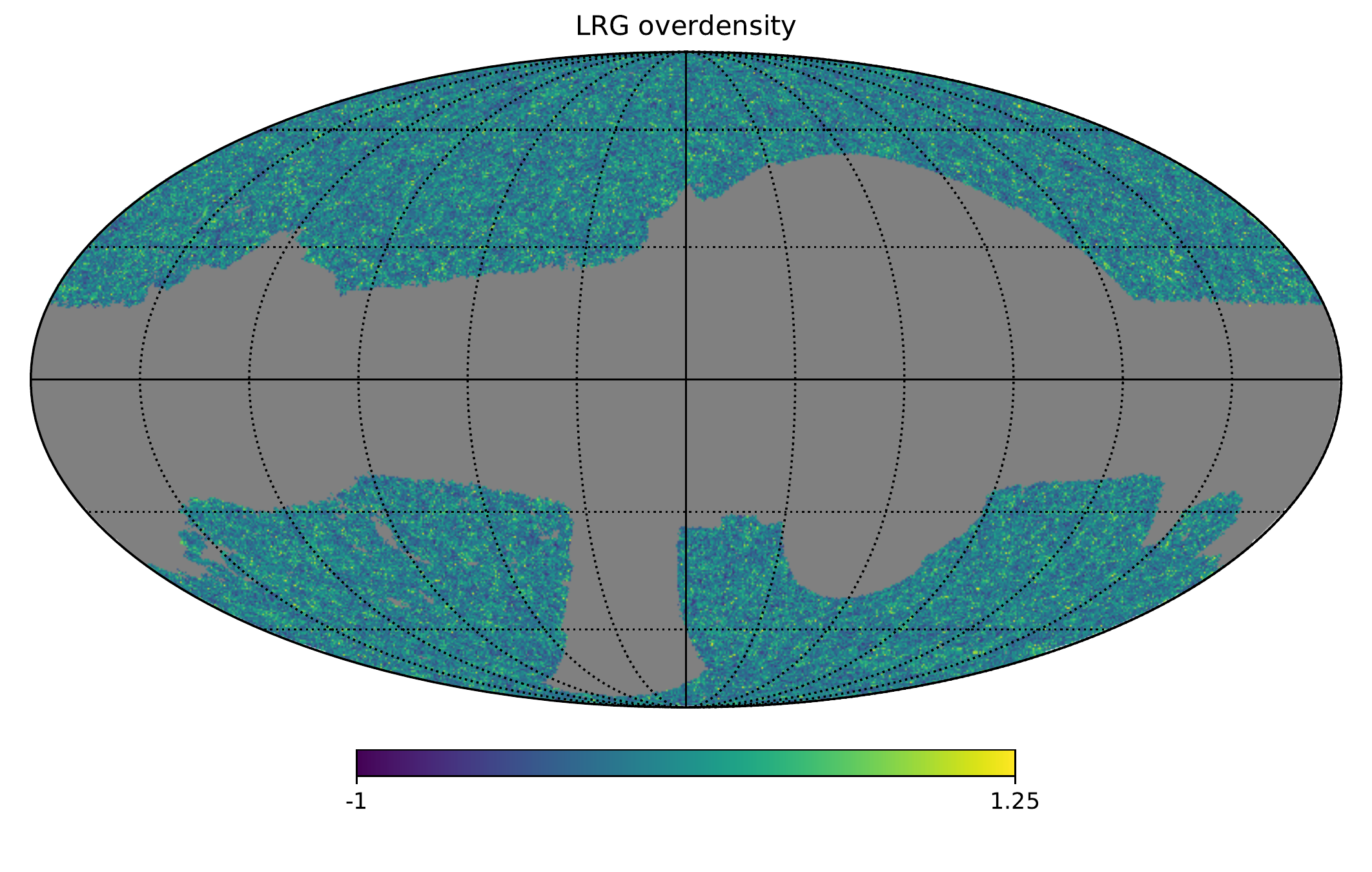}}
    \resizebox{0.8\columnwidth}{!}{\includegraphics{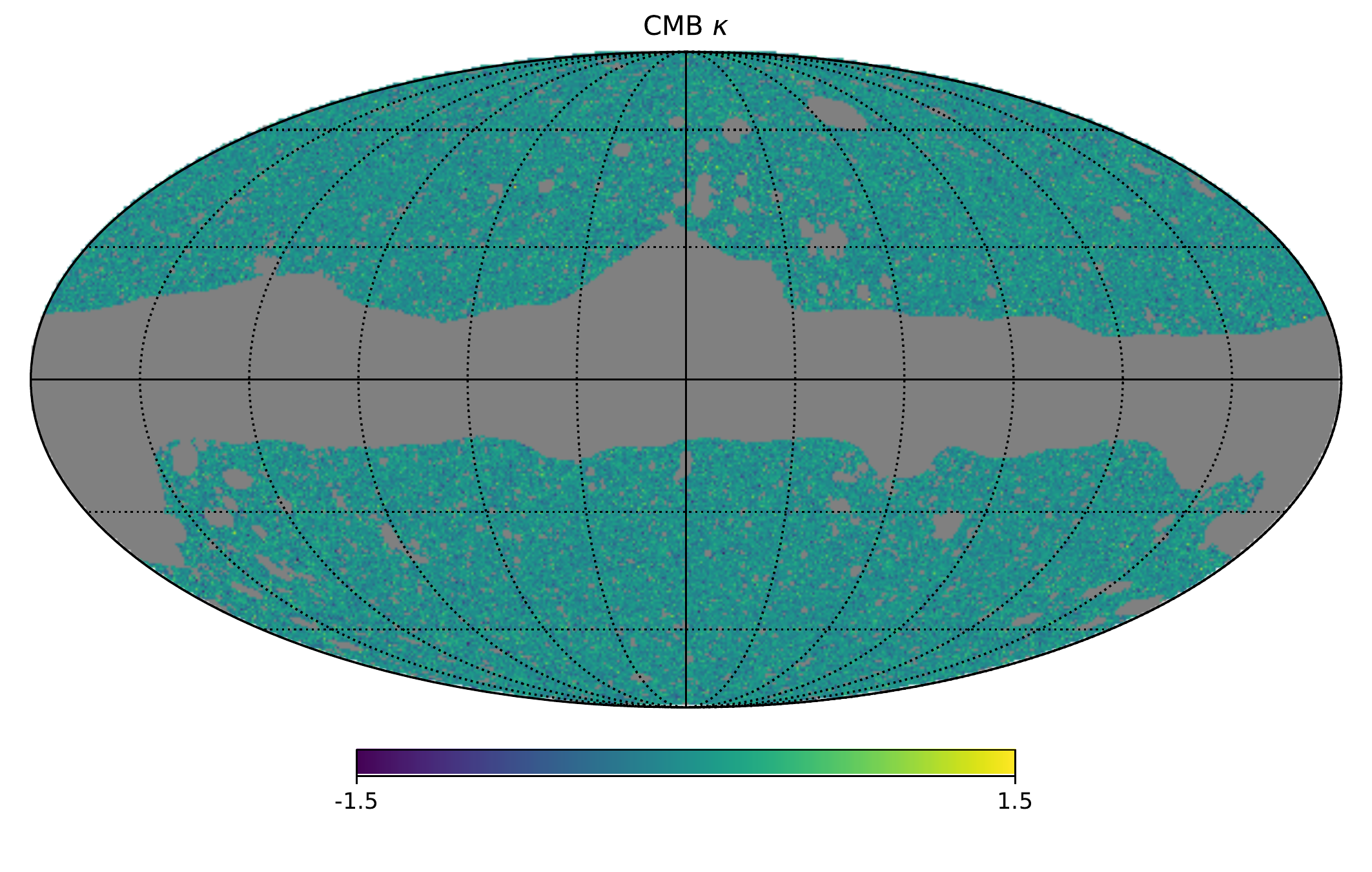}}
    \caption{The LRG overdensity for the full sample (top) and CMB-determined $\kappa$ (bottom), in Mollweide projection and galactic coordinates.  We follow the astronomical convention that east is left.  The grey region shows parts of the sky that are masked (see text).  The common, unmasked area is approximately 18,000 sq.deg.  The maps use the Healpix pixelization, with $N_{\rm side}=256$.}
    \label{fig:maps}
\end{figure}

Our galaxy samples are drawn from the imaging data \cite{Dey19} being used for target selection by the DESI spectroscopic redshift survey \cite{DESI}.  The samples are described in detail in ref.~\cite{PaperI} and some of the key features are summarized in Table \ref{tab:samples}.  Briefly they consist of DESI luminous red galaxies (LRGs) \cite{LRG_selection_paper}, selected from the imaging by applying a series of color cuts on extinction-corrected magnitudes in the $g$, $r$, $z$, and $W1$ bands, that have been divided into 4 redshift slices using photometric redshifts with 1.4--2.8 million galaxies per slice.  We refer to these photometric redshift bins as \texttt{pz\_bin} 1 through 4.  The combination of all 4 slices has 500 galaxies per square degree and covers 44\% of the sky ($\approx 18$K square degrees; Fig.~\ref{fig:maps}).  The photometric redshift cuts have been designed to give narrow redshift distributions with well-controlled tails at low and high redshift and modest overlap between samples as this simplifies the modeling of the cross-correlations (see ref.~\cite{PaperI} and later for further discussion).

In this paper we will focus on the `fiducial' sample of ref.~\cite{PaperI}.  This is drawn from the LRG sample for which DESI will obtain spectroscopic redshifts over the course of the survey.  While the legacy imaging surveys also provide deeper samples, with higher number density in the same redshift slices, we will be limited not by galaxy shot noise but rather by noise in the Planck $\kappa$ map and the limitations of our models.  For this reason we focus on the fiducial sample, with the goal of eventually combining these constraints with those from 3D clustering using DESI redshifts.  Other cross-correlation projects, including cross-correlations with higher fidelity CMB lensing maps, may benefit more from deeper samples which have lower shot noise while maintaining the narrow redshift distributions.  Since the DESI survey obtained deeper spectroscopy during its commissioning, the redshift distributions for these samples can still be spectroscopically calibrated.

We use the masks provided by ref.~\cite{PaperI}, which are chosen to eliminate bad data, avoid the galactic plane and regions of high extinction or high stellar density and to mask out bright stars.  While we make every effort to keep them to a minimum during catalog construction, the mask contains a large number of small holes and apodizing the mask (to smoothly connect the signal and masked regions) would lead to a non-trivial loss of area and invalidate some of the approximations we will use later to compute the covariance matrix (see e.g.\ ref.~\cite{Nicola20} for an extended and clear discussion).  Thus we do not apodize the galaxy mask.  Tests with Gaussian mock catalogs show that we can recover the bandpowers on the scales relevant to our analysis with 0.1\% precision whether or not we apodize the mask.  This is much smaller than our statistical errors.

\begin{figure}
    \centering
    \resizebox{\columnwidth}{!}{\includegraphics{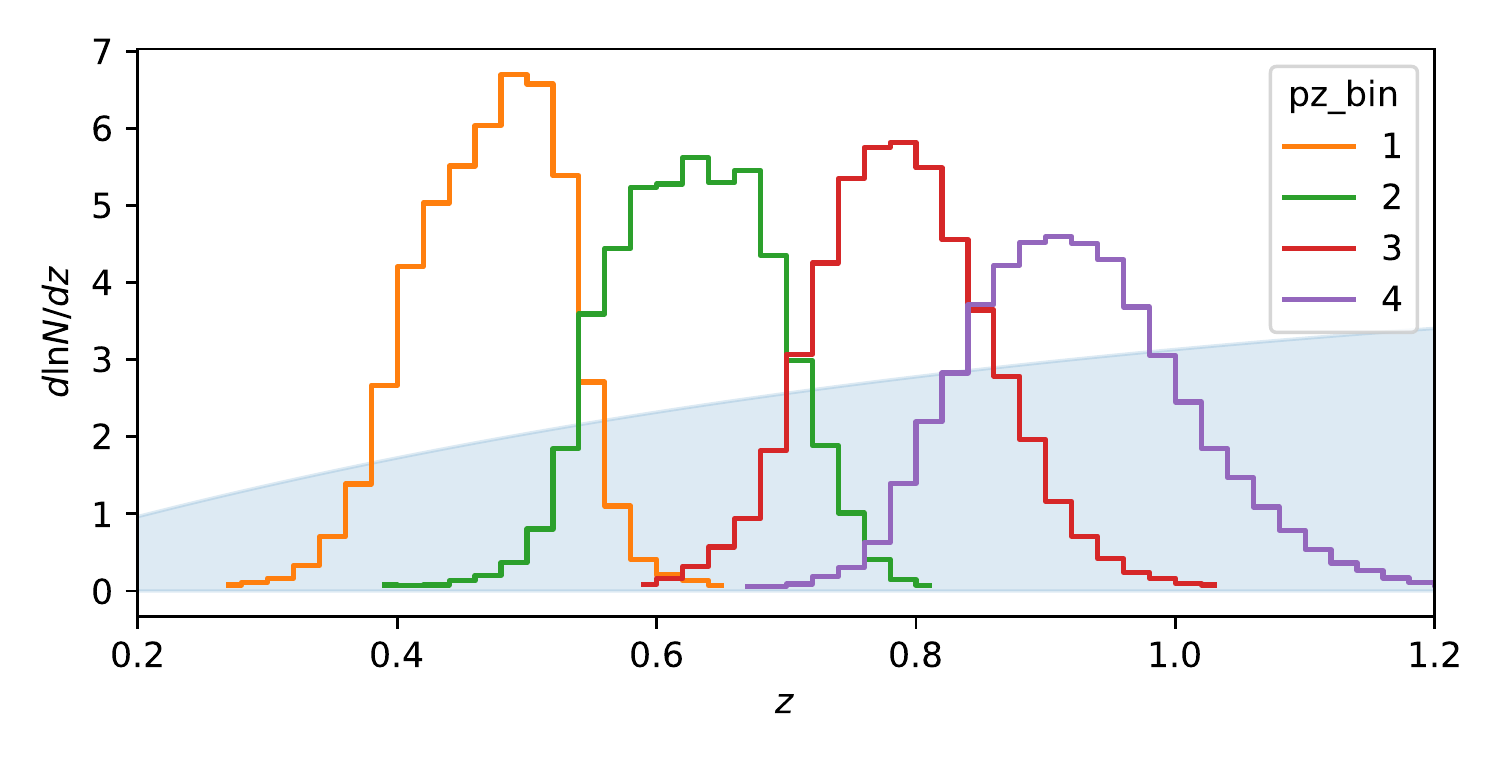}}
    \caption{The redshift distribution of our 4 photo-$z$-selected samples (Table \ref{tab:samples}), as determined by DESI spectroscopy.  The bins are relatively narrow and somewhat disjoint while maintaining a high angular number density, making them good for tomographic cross-correlation analyses.  The shaded band is proportional to the CMB lensing kernel.  }
    \label{fig:dndz}
\end{figure}

The maps are generated in the standard manner: the LRGs for each slice are binned into Healpix \cite{Gorski05} pixels at $N_{\rm side}=2048$ to form an ``LRG map''.  Random points are uniformly distributed within the footprint, and then passed through a series of masks that mirror the cuts applied to the data and include BRIGHT, GALAXY and CLUSTER masks\footnote{\url{https://www.legacysurvey.org/dr9/bitmasks/\#maskbits}} plus a custom bright star mask \cite{PaperI}.  We apply systematics weights to each random point to account for the effects of the largest systematics on the density field \cite{PaperI}.  The weights include corrections for trends with extinction, PSF size and depth in each band as discussed in detail in ref.~\cite{PaperI}.  These weights are applied to the random catalog rather than the LRGs themselves to reduce the impact of shot-noise from the small number of galaxies per pixel at $N_{\rm side}=2048$ (we have approximately 0.1 galaxies and 60 randoms per pixel per slice).  These weights are also used in the computation of the Poisson shot-noise level, which is listed in Table \ref{tab:samples}.  The LRG samples that we use are already very clean, and we find that these systematics weights have a very small effect on the measured clustering.  For all redshift bins including the weights removes some large-scale power associated with the very different imaging depths in the SGC compared to the NGC region (by $1-2\%$ at $\ell=50$; see ref.~\cite{PaperI} for more details and figures).  The weighted random counts in each Healpix pixel then form the ``random map''.  The overdensity field is defined as the ``LRG map'' divided by the ``random map'', normalized to mean density and mean subtracted.

Since the photometric systems between the northern and southern imaging differ, there is a possibility for systematics to affect one sample differently than the other.  We have checked that our results are stable (within errors) to selecting only galaxies from the northern vs.\ the southern hemisphere.  Our results are also stable to making a more stringent $E(B-V)$ cut, or to different treatments of the stellar mask.  Upon cross-correlating the galaxy maps with alternative extinction maps produced by the Planck collaboration \cite{Planck15X} we find only a $1-5\%$ correlation, suggesting this is not a major contribution to our systematic error budget.

The redshift distribution is obtained from spectroscopic redshifts obtained by DESI, as described in detail in ref.~\cite{PaperI}, and is shown in Fig.~\ref{fig:dndz}.  The LRG sample has a spectroscopic redshift completeness of 98\%. The 2\% LRGs without secure redshifts are mostly a result of low spectroscopic S/N, and we correct for this redshift incompleteness by up-weighting the fainter LRGs (that do have secure redshifts) in the $dN/dz$ estimation. Note the definition of the redshift bins through \texttt{pz\_bin} leads to quite compact redshift distributions with minimal tails, which greatly simplifies the modeling as we describe later. 

From the spectroscopic followup performed by DESI, we know that the stellar contamination is very low ($\ll 1\%$) for these samples and can be safely neglected.  There is a small contribution from active galactic nuclei, frequently an active nucleus within a luminous red galaxy, which have very similar $dN/dz$ and bias to our main sample and thus we shall not try to correct for this contamination.

Finally we require the slope of the galaxy number counts as a function of magnitude, $s_\mu$, in order the model the magnification bias of our samples (Table \ref{tab:samples} and Eq.~\ref{eqn:sdef}).  The slope is determined independently for each redshift bin by finite differences (shifting all magnitudes by $\delta m = 0.01$ and re-applying the selection) with two additional corrections.  First, since the LRG selection includes a fiber-flux cut we need to account for how the fiber-flux of each galaxy changes when it is magnified or demagnified.  This is estimated as a function of the shape parameters for each morphology type, as described in detail in ref.~\cite{PaperI}.  A second subtlety is that the redshift binning is done using random-forest-based photo-$z$s that use magnitude and galaxy size as features.  To account for the change in photo-$z$s under magnification we recompute the photo-$z$s with the shifted magnitudes and sizes and redo the redshift selection.  These two corrections change the slope from the naive finite difference estimate by $10-30\%$ (depending upon sample \cite{PaperI}).

\subsection{Planck lensing convergence}

We use the latest CMB lensing maps from the Planck 2018 release \cite{PlanckLens18} and their associated masks, downloaded from the Planck Legacy Archive.\footnote{PLA: \url{https://pla.esac.esa.int/}}
These maps are provided as spherical harmonic coefficients of the convergence, $\kappa_{\ell m}$, in HEALPix format \citep{Gorski05} and with $\ell_{\rm max} = 4096$.  In particular, for our fiducial analysis we use the minimum-variance (MV) estimate obtained from both temperature and polarization, based on the \texttt{SMICA} foreground-reduced CMB map.

We low-pass filter the $\kappa_{\ell m}$ to remove power above $\ell=2500$ before generating the convergence map in Healpix format \cite{Gorski05} at $N_{\rm side}=2048$ with the Healpix routine {\tt alm2map} (Fig.~\ref{fig:maps}).  We apodize the mask provided by the Planck team with a $30'$ ``C2'' filter\footnote{This applies a filtering $(1/2)[1-\cos \pi x]$ to all pixels near ($x<1$) a pixel in the mask with zero value (i.e.\ a masked pixel).  The variable $x\equiv\sqrt{(1-\cos\theta)/(1-\cos\theta_\star)}\approx \theta/\theta_\star$ with $\theta_\star$ the apodization scale.} using the NaMaster \cite{Alonso18} routine {\tt mask\_apodization}.  This makes the window functions we derive later more compact in harmonic space.

Since the MV reconstruction is dominated by temperature, residual galactic and extragalactic foregrounds may contaminate the signal. Extensive testing has been performed by the Planck team, indicating no significant problems at the current statistical level.  However, as a test, we repeated the analysis with a lensing reconstruction on \texttt{SMICA} foreground-reduced maps where the thermal tSZ effect \cite{SZ72} has been explicitly deprojected \cite{PlanckLens18}.  For the highest two redshift samples we found negligible impact from this change.  For the second bin the change was larger but not statistically significant.  However for the lowest redshift bin the cross-correlation with the tSZ-deprojected maps revealed $2\,\sigma$ excess power at $\ell\simeq 50$.  We believe this may be due to non-tSZ foregrounds in the lensing map (e.g.\ dust or CIB) correlating with systematics in the LRG map.  It is known that SZ-deprojection increases the impact of non-SZ foreground on the $\kappa$ map \cite{Sailer:2021vpm}.  We also know that the lowest redshift bin could be more affected by galactic dust or extinction due to its use of the $g$-band to reject low $z$ galaxies. The combination could plausibly explain the excess power at low $\ell$.  Though we have no evidence that the cross-correlation with the fiducial lensing maps is similarly contaminated, this may suggest the cross-correlations with the LRGs in the first redshift bin are slightly less reliable, at least on large scales.  Since most extinction maps are correlated with large-scale structure due to contamination from the CIB \cite{Lenz:2017djx,Yahata:2006mf}, one direction for future work could be a study of the correlation with CIB-free extinction or reddening maps such as those derived from galactic H{\sc i} emission \cite{Lenz:2017djx}. We also note\footnote{We thank the referee for emphasizing this point.} that the provided CMB lensing mask removes regions around high $S/N$ tSZ sources to avoid contamination.  These are mostly low-redshift sources that could potentially correlate with the lowest redshift bin \cite{Lembo:2021kxc}.  We shall leave further investigation of these issues to future work.

\subsection{Angular power spectra and covariance}
\label{sec:spectra}

\begin{figure}
    \centering
    \resizebox{\columnwidth}{!}{\includegraphics{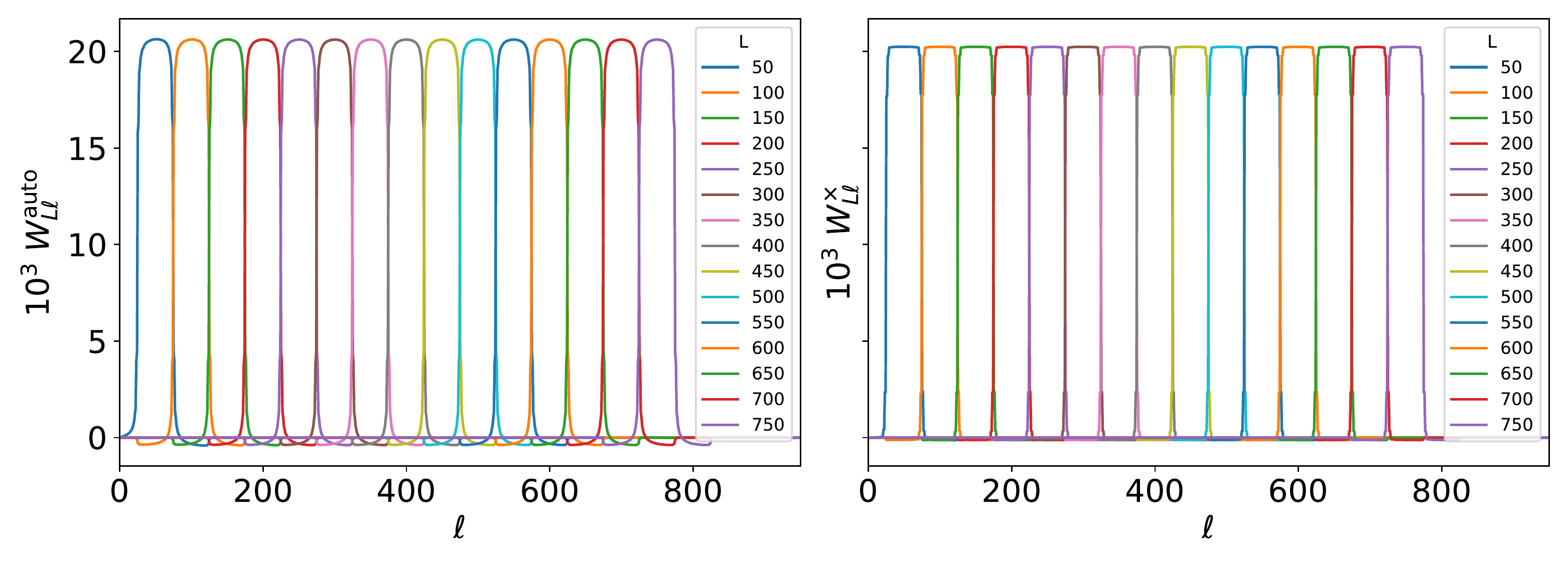}}
    \caption{The bandpower window functions for the auto- (left) and cross- (right) spectra defined such that $\langle C_L^{\rm obs}\rangle = \sum_\ell W_{L\ell} C_\ell^{\rm thy}$.  Since we bin $C_L^{\rm obs}$ with $\Delta\ell=50$ we expect $W_{L\ell}$ to be approximately top-hats of height $0.02$ and width $50$.  The effect of masking modifies this expectation somewhat, with the largest impact on the auto-correlation window function.}
    \label{fig:bandpower_windows}
\end{figure}

The estimation of angular power spectra from sky maps is now routine (see refs.~\cite{Nicola20,Marques20,Krolewski20,Singh20,Darwish21,Hang21a,Kitanidis21,Zhang21a,Robertson21,Shekhar21,Garcia21,DES_Y3_BAO,Sun21} for some examples within just the last year).  We use the pseudo-$C_\ell$ method \cite{Hivon02} as implemented within the \texttt{NaMaster} package \cite{Alonso18} to estimate our angular power spectra.  As a reminder, this approach uses fast spherical harmonic transforms to expand a field, $X(\hat{n})$ into spherical harmonics, $Y_{\ell m}$, as
\begin{equation}
    \tilde{a}_{\ell m} = \int_{4\pi}d\Omega \ X(\hat{n}) \ W(\hat{n}) \ Y^{*}_{\ell m}(\hat{n})
\end{equation}
where $W(\hat{n})$ is the mask.  The (pseudo) angular power spectrum, $\tilde{C}_{\ell}$, is then an average over $m$-modes,
\begin{equation}
    \tilde{C}_{\ell} = \frac{1}{2\ell + 1}\sum_{m=-\ell}^{\ell} |\tilde{a}_{\ell m}|^2
\end{equation}
The $\tilde{C}_\ell$ are related to the true spectrum via a convolution with the mask such that their ensemble averages are related by $\langle \tilde{C}_{\ell} \rangle = \sum_{\ell^{\prime}} M_{\ell \ell^{\prime}} \langle C_{\ell^{\prime}} \rangle$ where the mode-mode coupling matrix $M$ can be determined purely from the mask \cite{Hivon02}.
This $\ell$-by-$\ell$ matrix is generally singular in the case of large sky cuts. In order to perform matrix inversion, a common method is to use a set of discrete bandpower bins $L$ and assume the angular power spectrum is a step-wise function in each bin.  While not strictly necessary, the $L$-by-$L$ mode-mode coupling matrix is typically inverted and applied to extract the binned angular power spectrum from the binned pseudo angular power spectrum ($\langle C_{L} \rangle = \sum_{L^{\prime}} M_{L L^{\prime}}^{-1} \langle \tilde{C}_{L^{\prime}} \rangle$).  The final result is a linear relationship between the binned power spectrum, $C_L$, and the underlying $C_\ell$ that holds for expectation values. 
We use the \texttt{compute\_full\_master} method in \texttt{NaMaster} \cite{Alonso18} to calculate the binned power spectra and the bandpower window functions relating them to the underlying theory: $\langle C_L\rangle = \sum_\ell W_{L\ell} C_\ell$.
We choose a conservative binning scheme with linearly spaced bins of size $\Delta\ell = 50$ starting from $\ell_{\rm min} = 25$.  The bin width is larger than expected correlations between modes induced by the survey masks, while being narrow enough to preserve the structure in our angular spectra. To avoid power leakage near the edge of the measured range we perform the computation to $\ell = 6000$, and simply discard the bins beyond some $\ell_{\rm max}$ \cite{Krolewski20}.  The associated $W_{L\ell}$ for the auto- and cross-power spectra are shown in Fig.~\ref{fig:bandpower_windows} for the bandpowers up to $L=750$.

While it is a small effect on the scales of interest here, we correct our power spectra for the effects of the pixel window function (obtained from the Healpix routine \texttt{pixwin}).  The cross-spectrum between $\kappa$ and the galaxy density is divided by one power of the pixel window function while the shot-noise-subtracted auto-spectrum is corrected by two powers.  The shot-noise itself receives no correction.  These choices were checked against Gaussian and Poisson sky simulations.

\begin{figure}
    \centering
    \resizebox{\columnwidth}{!}{\includegraphics{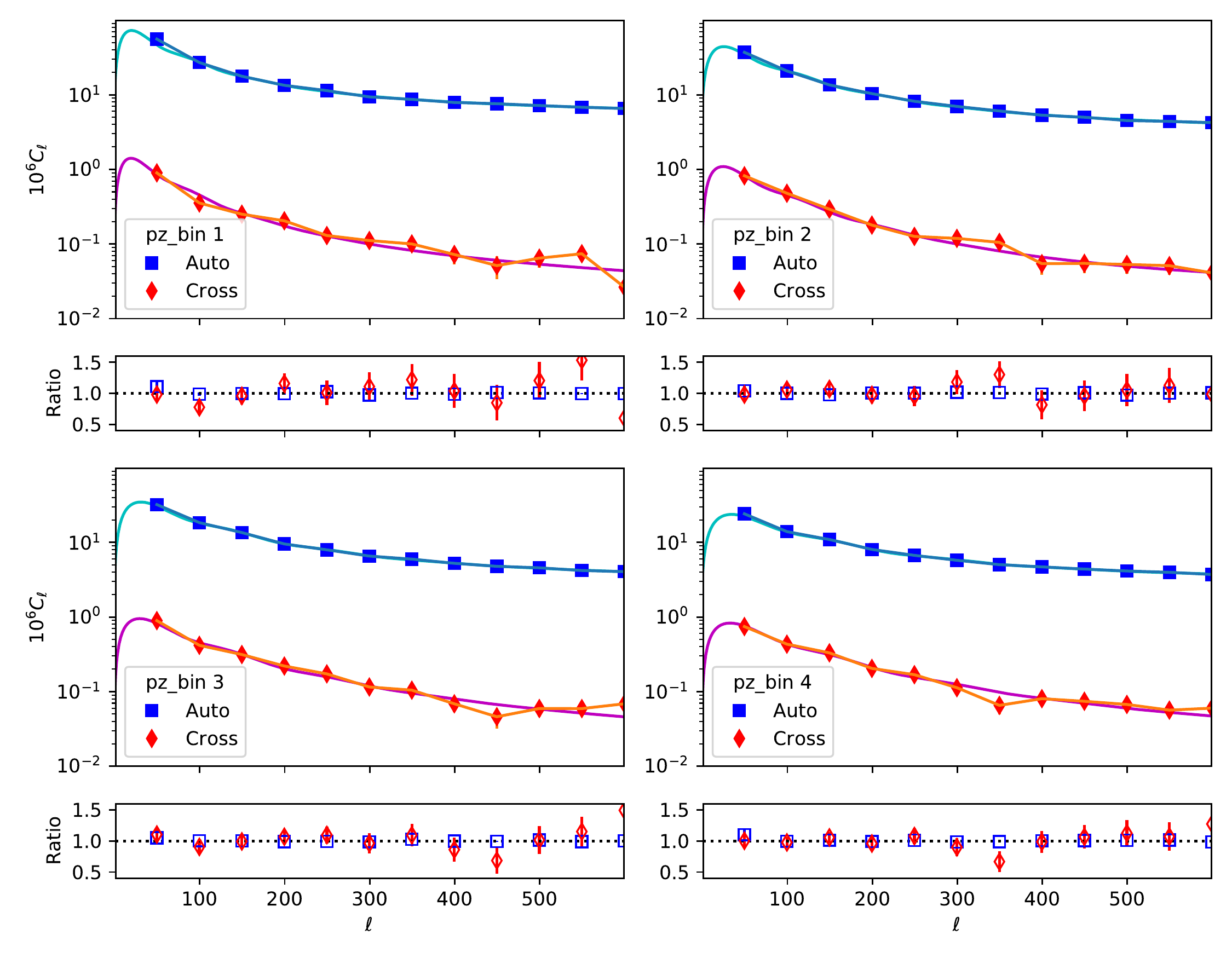}}
    \caption{The angular pseudo-spectra ($C_\ell^{gg}$ in blue and $C_\ell^{g\kappa}$ in red) for each of our 4 samples in bins of $\Delta\ell=50$.  The points, with errors, show the measurements while the solid lines show the fit used to compute the analytic covariance matrix.  The second and fourth rows show the ratio of the data to the fit, so that it is easier to see the errors (which are very small, especially for the autocorrelation).}
    \label{fig:bandpowers}
\end{figure}

\begin{figure}
    \centering
    \resizebox{\columnwidth}{!}{\includegraphics{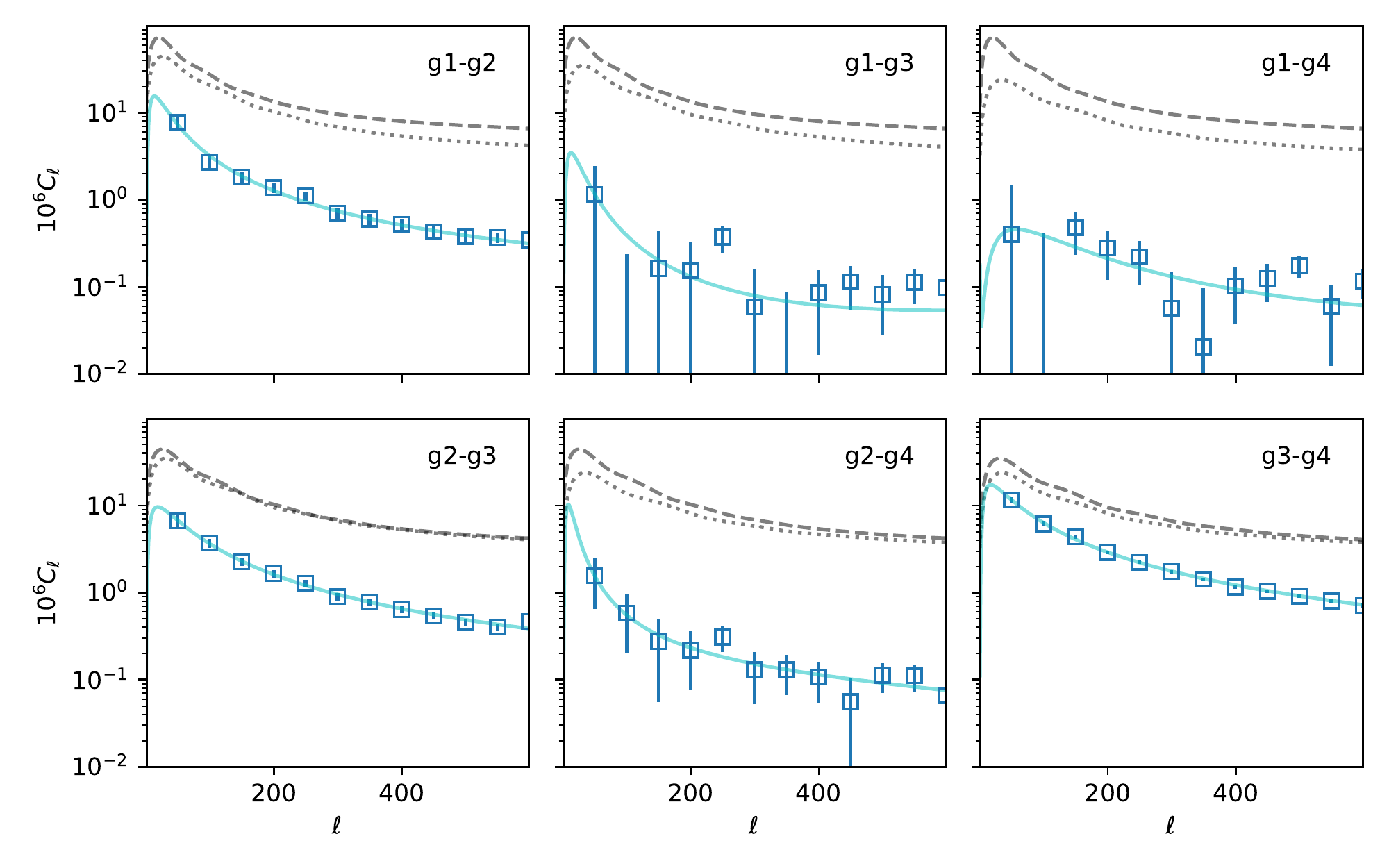}}
    \caption{The angular pseudo-cross-spectra ($C_\ell$) for each of our 4 galaxy samples cross-correlated with each of the others.  In each panel the dashed line shows the fit to the auto-spectrum of the lower redshift sample while the dotted line shows the fit to the higher redshift sample for comparison.  The points show the measured cross-correlation while the light cyan line shows a polynomial fit used in computing the covariance matrix.}
    \label{fig:galaxy_cross_power}
\end{figure}

Fig.~\ref{fig:bandpowers} shows the measured bandpowers, $C_\ell^{gg}$ and $C_\ell^{\kappa g}$, for each of our samples.  The error bars are the diagonal of the covariance matrix, but the individual bandpowers are largely decorrelated with bins of width $\Delta\ell=50$ and our sky coverage.  Fig.~\ref{fig:galaxy_cross_power} shows the cross-spectra between galaxies in different slices, which we use to compute the covariance matrix of the joint samples.  Since the photo-$z$ bins are relatively compact and have little overlap the cross-power is a small fraction of the auto-power in each slice.

\begin{figure}
    \centering
    \resizebox{\columnwidth}{!}{\includegraphics{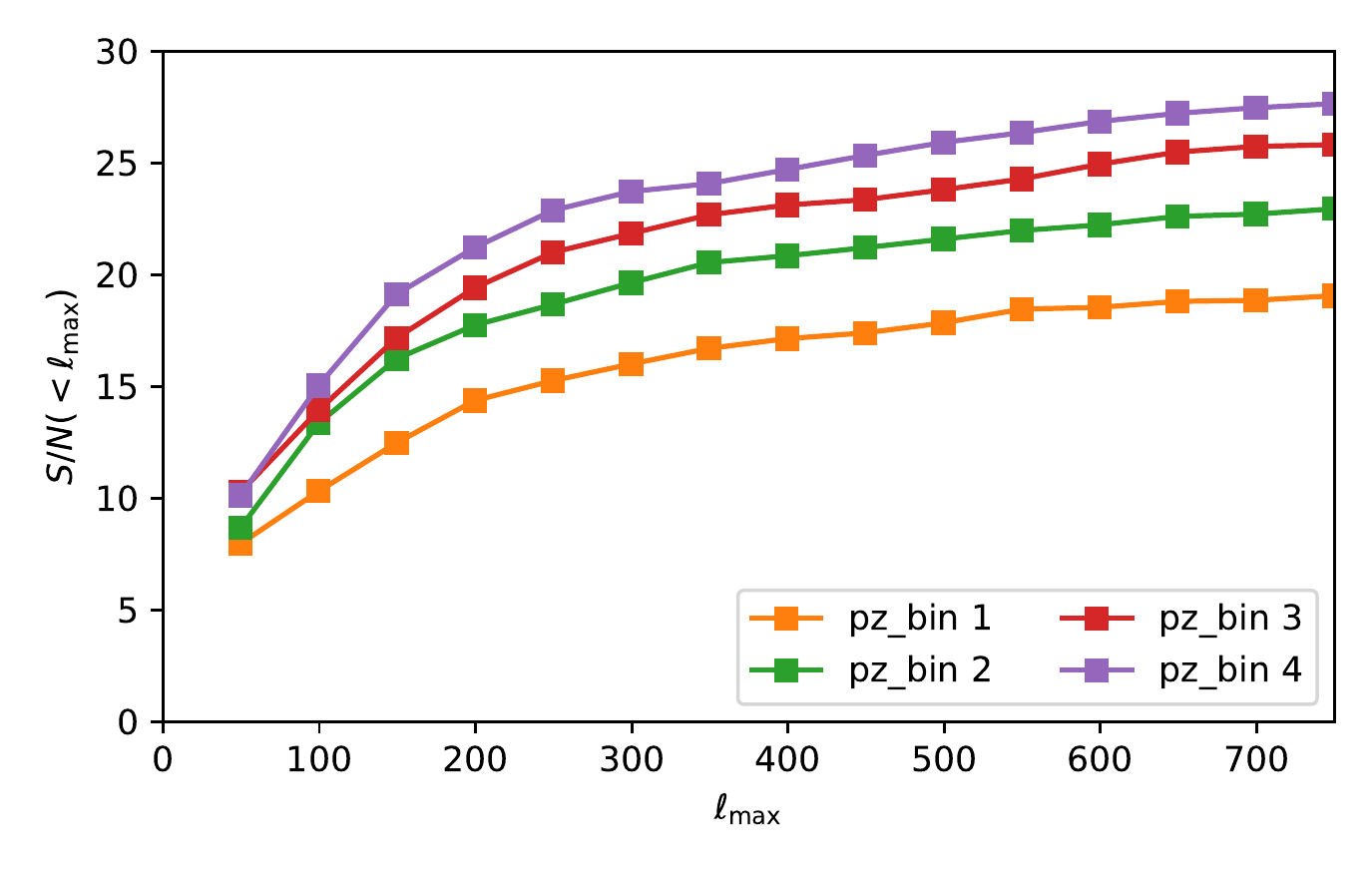}}
    \caption{The (cumulative) signal to noise ratio for $C_\ell^{g\kappa}$ for our 4 samples.  The signal-to-noise ratio of $C_\ell^{gg}$ is much higher than $C_\ell^{\kappa g}$, thus the latter dominates the error when determining cosmological parameters.  While the samples are not independent, so the SNRs cannot be simply added, this figure serves to indicate the range of scales and which samples dominate our constraints.}
    \label{fig:snr}
\end{figure}

\begin{figure}
    \centering
    \resizebox{\columnwidth}{!}{\includegraphics{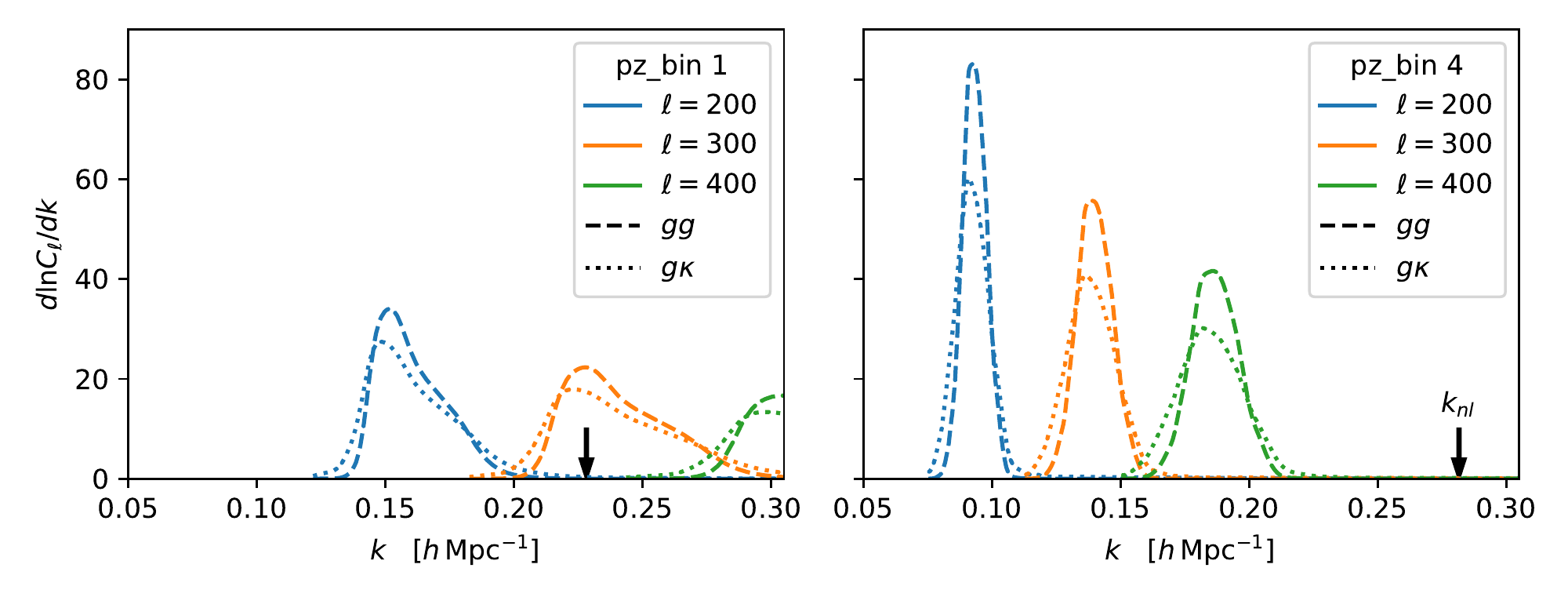}}
    \caption{The fractional contributions to $C_\ell^{gg}$ (dashed) and $C_\ell^{g\kappa}$ (dotted) for three different multipoles ($\ell=200$, 300 and 400; blue, orange and green) for the lowest (left; $z\approx 0.5$) and highest (right; $z\approx 0.9$) redshift samples.  We use the best-fitting \texttt{Anzu} model for the signal model.  Within the Limber approximation the contributions are given by Eq.~(\ref{eqn:ClLimber}), with a given multipole probing scales $k\approx \ell/\chi_0$ if $\chi_0$ is the characteristic distance to the sample.  The black, downward-pointing arrows show the non-linear scale, $k_{\rm nl}$, defined as the inverse of the mean Zeldovich displacement (i.e.\ the same convention as ref.~\cite{Sailer21}).  The scales where our S/N peaks are in the linear or quasi-linear regime, with negligible contributions from baryonic effects, simplifying the analysis. }
    \label{fig:kernel}
\end{figure}

Our signal comes predominantly from large scales (Figs.~\ref{fig:snr}, \ref{fig:kernel}) where the approximation of a Gaussian (disconnected) covariance matrix holds.  We compute this matrix with the spectra of the best-fitting theoretical model from an earlier iteration using the NaMaster function  \texttt{gaussian\_covariance}, including both the auto-correlations of $C_\ell^{gg}$ and $C_\ell^{\kappa g}$ and their covariance.  The signal and noise of the $\kappa$ autospectrum are taken from the Planck data release.  We checked that the simpler approximation in \S 3 of ref.~\cite{Hivon02}, which involves the $2^{\rm nd}$ and $4^{\rm th}$ moments of the window function and $f_{\rm sky}$, also works well, with the full expression giving slightly stronger bin-bin correlations but quite similar $\chi^2$ values in our fits (i.e.\ differences much less that $\sqrt{2\,N_{\rm dof}}$).  We also checked our covariance matrix against one produced by Monte-Carlo sampling 2048 Gaussian realizations, with the same masking, binning, etc.  The agreement was also good.  Denoting $C_{\rm sim}$ as the Monte-Carlo derived covariance and $C_{\rm fid}$ as our fiducial (analytically derived) one ${\rm tr}\left[C_{\rm fid}\ \Delta\Omega\right]$, with $\Delta\Omega = C_{\rm fid}^{-1}-C_{\rm sim}^{-1}$, gives the `typical' change in $\chi^2$ induced by using the different covariance matrices.  For our data in the most constraining slice (\texttt{pz\_bin} 4) ${\rm tr}\left[C_{\rm fid}\ \Delta\Omega\right]\approx 0.2$ for the 8 data points in the galaxy auto-spectrum below $\ell_{\rm max}=400$ most affected by masking, much smaller than $\sqrt{2\,N_{\rm dof}}$.  We note that these approximations become less accurate as the number of holes in the mask increases, as this leads to a broader window and violation of some of the approximations made in NaMaster \cite{Nicola20}.  It is for this reason that we have kept the number of holes as small as practically possible \cite{PaperI}.  The agreement between the Monte-Carlo and NaMaster-produced covariance was not as good for $C_\ell^{\kappa g}$, with the Monte-Carlo predicting slightly smaller variance at low $\ell$.  However, the variance at higher $\ell$ and the structure of the correlation matrix was very similar (up to noise) so we have chosen to use the NaMaster-produced covariance as the conservative choice.  As a cross check if we use the Monte-Carlo covariance instead of our default covariance when fitting $\Lambda$CDM models to each \texttt{pz\_bin} sample the best-fit parameters change by $<0.2\,\sigma$.  For all of our approximations the covariance matrices are strongly diagonally dominant, with the first off-diagonal terms typically at the percent level and second off-diagonals even lower.  This is because our bin width, $\Delta\ell=50$, and our sky coverage are both simultaneously large.

When we fit multiple samples at once we do not include the galaxy cross-spectra (which are quite small; Fig.~\ref{fig:galaxy_cross_power}) in the data vector but we do include their contributions to the covariance (based on each of the $C_\ell^{g_ig_j}$ and $C_\ell^{g_i\kappa}$).  The inclusion of the galaxy cross-spectra as data would be helpful in calibrating $dN/dz$ \cite{Schaan:2020qox}, but since we have a spectroscopically calibrated $dN/dz$ they contain little additional information.

\section{Modeling}
\label{sec:modeling}

\subsection{Spectra}

The galaxy overdensity, $\delta_g$, and CMB lensing convergence, $\kappa$, are both projections of 3D density fields.  Within the Limber approximation \citep{Limber53} the angular cross-spectrum between two such fields $X$ and $Y$ is given by\footnote{The largest correction to the Limber approximation, arising at low $\ell$, comes from redshift-space distortions (RSD) in the galaxy auto-spectrum \cite{Fisher94,Padmanabhan07}.  These effects are smaller than our errors for the bandpowers we consider except for the first bin where for some samples it is comparable.  Rather than include RSD in our predictions, we add a ``theory error'' to the first bin conservatively larger than the difference between RSD and no-RSD predictions of our fiducial model.} an integral over comoving distance $\chi$
\begin{equation}
C_{\ell}^{XY} = \int d\chi \ \frac{W^X(\chi) W^Y(\chi)}{\chi^2} P_{XY}{\Big (}k = (\ell + 1/2)/\chi; z{\Big )}
\label{eqn:ClLimber}
\end{equation}
where the projection kernels for galaxy overdensity and CMB lensing convergence are, respectively,
\begin{gather}
\begin{aligned}
    W^{g}(\chi) &= H(z(\chi))\phi(z(\chi)) + W^\mu(\chi) \\
    W^{\kappa}(\chi) &= \frac{3}{2} \Omega_{m0}H_0^2 (1+z) \frac{\chi(\chi_{\rm *}-\chi)}{\chi_{\rm *}} 
\end{aligned}
\end{gather}
with $\chi_{\rm *} = \chi(z_{\rm *}{\approx}1100) \approx 9400$ $h^{-1}$Mpc the distance to the surface of last scattering and $\phi(z)$ the normalized redshift distribution of the galaxy sample.  The $W^\mu$ term in $W^g$ accounts for magnification bias and, to first order, is given by
\begin{align}
    W^{\mu}(\chi) &= (5s_\mu-2)\ \frac{3}{2} \Omega_{m0}H_0^2 (1+z) \int_z^{z_{\star}} dz^{\prime}\ g(z^{\prime}) \\
    g(z^{\prime}) &= \frac{\chi(z)(\chi(z^{\prime})-\chi(z))}{\chi(z^{\prime})} \phi(z^{\prime})
\end{align}
with $z$ implicitly a function of $\chi$ and $s_\mu$ the slope of the cumulative magnitude function, i.e.\ the response of the number density of the sample to a multiplicative change in brightness at the limiting magnitude of the survey:
\begin{align}
  s_\mu \equiv \left. \frac{d\log_{10}n(m<m_{\rm lim})}{dm}\right|_{m=m_{\rm lim}}
  \quad .
\label{eqn:sdef}
\end{align}
Due to the relatively narrow $dN/dz$ of our samples the magnification correction is small for the auto-spectrum.  It is larger for the cross-spectrum, but even there amounts to only 5\% of the signal even for the worst case (the highest redshift bin), so we are quite insensitive to uncertainties in this term.  Since the magnification term probes smaller scales than the other contributions to the auto- and cross-spectra but high accuracy is not a requirement we use the \texttt{HaloFit} fitting function for $P_{mm}$ (\cite{Smith03,Takahashi12} as implemented in \texttt{CAMB} \cite{Lewis00,Howlett12}) for the magnification contributions.  We have checked that using perturbation theory with the lowest order counterterm instead leads to indistinguishable posterior distributions.

To build intuition we note that for a thin shell of galaxies at distance $\chi_0$ of width $\Delta\chi$, if the other parts of the integrands are changing slowly we have
\begin{equation}
    C_\ell^{gg} \approx \mathcal{V}^{-1}P(k=\ell/\chi_0) \quad , \quad
    C_\ell^{g\kappa} \approx  W^\kappa(\chi_0)\chi_0^{-2}P(k=\ell/\chi_0) 
\end{equation}
where $\mathcal{V}=\chi_0^2\,\Delta\chi$ is the volume per steradian and we have neglected the magnification terms for simplicity.  Note the galaxy auto-spectrum increases as we decrease the width of the shell (and there is less ``washing out'' of the clustering signal) while the cross-spectrum amplitude is independent of $\Delta\chi$.  This implies the cross-correlation coefficient between $\kappa$ and the galaxy overdensity decreases for thinner shells, which probe a smaller fraction of the total $\kappa$ signal.

Since the Planck lensing map has high signal-to-noise predominantly at low $\ell$ (Fig.~\ref{fig:snr}), our measurement is dominated by large scales.  The relatively compact nature of $dN/dz$ (Fig.~\ref{fig:dndz}) means each $\ell$ mode probes only a narrow range of physical scales, $k$, and these modes will in general be quasi-linear at the redshifts of relevance (Fig.~\ref{fig:kernel}).  This, plus the spectroscopically determined $dN/dz$ and the well-known redshift of the CMB (which is always well behind and well separated from our lenses) makes our measurement robust to many common modeling uncertainties.  Figures \ref{fig:snr} and \ref{fig:kernel} also nicely illustrate the gains from moving to higher redshift galaxy samples.  The increase in the CMB lensing kernel from $z\sim 0$ to $z\sim 2$, the shift to probing more modes at higher $\ell$ for a fixed $k$ and the larger $k$ range for which the models are accurate at higher redshift all contribute to making the higher redshift samples more powerful probes.

\subsection{Power spectrum models}
\label{sec:pk_models}

Key to interpreting our measurements are  models for the galaxy-galaxy and galaxy-matter power spectra accurate on linear and quasi-linear scales.  Our cosmological results are based on a perturbative model (\S\ref{sec:clpt}), though we use an N-body based emulator (\S\ref{sec:anzu}) to validate some of our results.  We next describe each model, before discussing our parameter priors and some technical details of the implementation.

\subsubsection{Lagrangian perturbation theory}
\label{sec:clpt}

Our fiducial model is based on convolution Lagrangian effective field theory (CLEFT; \cite{Vlah16} and the references contained therein), as implemented in the \texttt{velocileptors}\footnote{Code available at \url{https://github.com/sfschen/velocileptors}} code \cite{Chen20,Chen21a}.  This is the same formulation and code as used in ref.~\cite{Kitanidis21}.  Under this formalism, the galaxy-galaxy and galaxy-matter power spectra are:
\begin{align}
    P_{\rm gg} &= \left(1-\frac{\alpha_{a}k^2}{2}\right)P_{\rm Z} + P_{\rm 1-loop} + b_1 P_{\rm b_1} + b_2 P_{\rm b_2} + b_1 b_2 P_{\rm b_1 b_2}  + b_1^2 P_{\rm b_1^2} + b_2^2 P_{\rm b_2^2} \\
    P_{\rm gm} &= \left(1-\frac{\alpha_{\times}k^2}{2}\right)P_{\rm Z} + P_{\rm 1-loop} + \frac{b_1}{2}P_{\rm b_1} + \frac{b_2}{2}P_{\rm b_2}
\label{eqn:lpt}
\end{align}
where we have dropped the terms corresponding to shear bias as we find they mainly affect small scales and are somewhat degenerate with the $\alpha_i$ \cite{Modi17}.  Here, $P_{\rm Z}$ and $P_{\rm 1-loop}$ are the Zeldovich and 1-loop dark matter contributions, $b_1$ and $b_2$ are the Lagrangian bias parameters \cite{Matsubara08} for the galaxy sample, and $\alpha_{\times}$ and $\alpha_a$ are free parameters encapsulating the small-scale physics not modeled by perturbation theory (see e.g.\ ref.~\cite{Vlah15} for further discussion).  Each term can be written as an integral of one or two powers of the linear theory power spectrum, times an analytically known kernel function.  Explicit formulae for the power spectrum contributions can be found in the above-mentioned references and are collected in Appendix \ref{app:lpt}.

To reduce model expense and avoid needing to model redshift-dependent bias, we evaluate the power spectrum terms at a single\footnote{Since the magnification terms depend upon a wider redshift range, we include the expected linear growth of $P_{mm}(k)$ in the kernel.  This is only an approximation on small scales, but the magnification is a small correction and we probe primarily large scales so the approximation is quite good.} effective redshift,
\begin{equation}
    z_{\rm eff}^{XY} = \frac{\int d\chi \ z \ W^{X}(\chi)W^{Y}(\chi)/\chi^2}{\int d\chi \ W^{X}(\chi)W^{Y}(\chi)/\chi^2} \quad .
\label{eqn:zeff}
\end{equation}
These redshifts are listed in Table \ref{tab:samples} for each sample for the fiducial Planck cosmology, though in our inferences we recompute $z_{\rm eff}$ and $P(k,z_{\rm eff})$ for each cosmology being tested. Given the narrow redshift distribution and almost passive bias evolution of our galaxy sample, the $z_{\rm eff}$ approximation does not affect the $C_{\ell}$'s significantly \cite{Modi17} and is very weakly cosmology dependent.  We validate this approximation using simulations later.

It is by now well established that halos and galaxies can be well approximated as biased tracers of the dark matter and baryon fluids only \cite{Villaescusa-Navarro2013,Castorina2013,Castorina2015,Villaescusa-Navarro2017,LoVerde2014,Munoz2018,Fidler2018}.  For this reason we use $P_{cb}$, the dark matter plus baryon spectrum as computed by the Boltzmann code CAMB \cite{Lewis00,Howlett12}, as input to \texttt{velocileptors} for $P_{gg}$ and $P_{gm}$.  We use the full matter power spectrum for $P_{mm}$. This substitution assumes the Green's function associated with higher order perturbative solutions is very well approximated by the Einstein-de-Sitter one, implying the structure of the loops remains unchanged. This is an excellent approximation for the small fiducial value of neutrino masses considered in this work \cite{Aviles:2020cax}. In principle one needs to consider the scale-dependence of the growth rate, which we shall neglect for the same reason mentioned above.

\begin{figure}
    \centering
    \resizebox{\columnwidth}{!}{\includegraphics{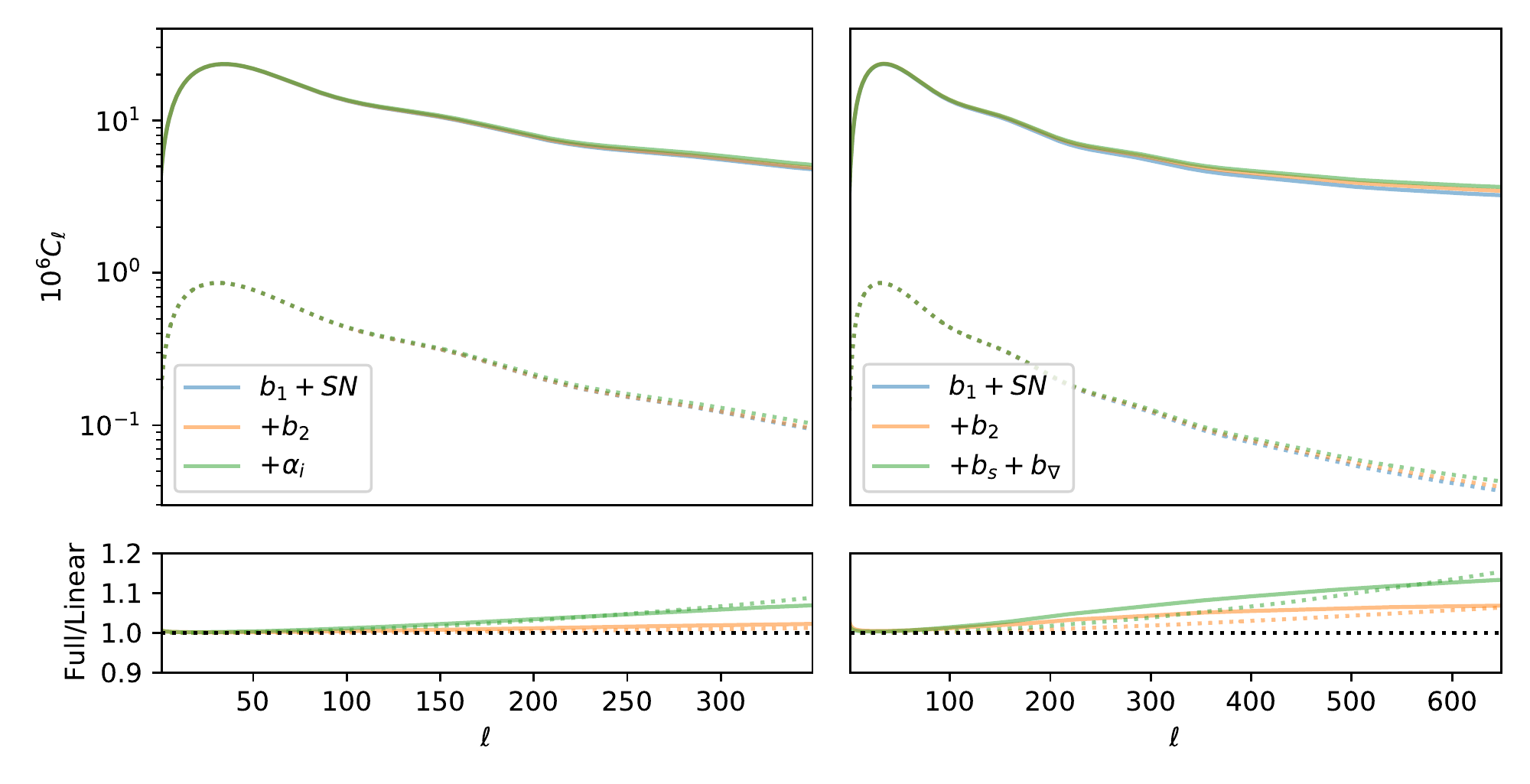}}
    \caption{Model predictions for $C_\ell^{gg}$ (solid) and $C_\ell^{\kappa g}$ (dotted) for the \texttt{pz\_bin}=4 sample as we successively add terms to the model.  The left panels show the Lagrangian PT results with the blue line being a base model with just $b_1$ ($=1.4$) and shot noise ($=5490$), the orange line adding $b_2\simeq 0.5$ and the green line including the counter terms ($\alpha_a=22$, $\alpha_\times=8$).  The right panels (plotted to higher $\ell_{\rm max}$) show the results for \texttt{Anzu}, again with the blue line including just $b_1$ ($=1.4$) and shot noise ($=5300$), the orange line adding $b_2\simeq 1$ and the green line adding $b_s\simeq -0.1$ and $b_\nabla\simeq -2.6$.  In each case we vary the bias parameters at fixed cosmology ($\Omega_m=0.31$, $h=0.68$, $\sigma_8=0.77$).  Note that on the scales plotted the lowest order results dominate over the effects of scale-dependent bias (and counterterms) for both the auto- and cross-spectra.  However, the observational errors are sufficiently small that the corrections are not insignificant.}
    \label{fig:components}
\end{figure}

Fig.~\ref{fig:components} shows $C_\ell^{gg}$ and $C_\ell^{\kappa g}$ for \texttt{pz\_bin}=4 as we successively add $b_2$ and counter terms to the model with just $b_1$ and shot noise.  The models for the other redshifts look qualitatively similar.  For illustrative purposes we vary the parameters for this sample at fixed cosmology ($\Omega_m=0.31$, $h=0.68$, $\sigma_8=0.77$).  Note that the scales we will probe are dominated by the linear contributions, however we find that using a ``minimal model'' with scale-independent bias and no counter terms can lead to shifts in parameters of $\mathcal{O}(1\,\sigma)$ depending upon the redshift.

\subsubsection{Lagrangian bias expansion: Anzu}
\label{sec:anzu}

The second model we employ is \texttt{Anzu}\footnote{Code available at \url{https://github.com/kokron/anzu}.} \cite{Kokron21a}, an emulator based on the Aemulus simulations \cite{deRose18} that implements the Lagrangian bias prescription with component spectra measured from N-body simulations \cite{Modi20}.  \texttt{Anzu} provides $P_{gg}$ and $P_{gm}$ as a function of four bias parameters (the same\footnote{The definition of $b_2$ employed in \texttt{Anzu} differs from that in \texttt{velocileptors} by a factor of two.  We have standardized to the \texttt{velocileptors} definition.} $b_1$, $b_2$ as above plus a shear bias, $b_s$, and a derivative bias, $b_\nabla$) for a range of $\Lambda$CDM cosmologies by interpolating from a grid.  \texttt{Anzu} does not (yet) include massive neutrinos so for this case we consider pure $\Lambda$CDM cosmologies.  In order to better match the shape of the linear power spectrum including massive neutrinos we reduce $n_s$ by 0.005 for \texttt{Anzu}.  This is a small change, and within the current uncertainty on $n_s$, but achieves a better match to the broad-band power between the models.

Fig.~\ref{fig:components} shows $C_\ell^{gg}$ and $C_\ell^{\kappa g}$ for \texttt{pz\_bin}=4 as we successively add the higher order biases to the model with just $b_1$ and shot noise.  For illustrative purposes we choose best-fit parameters for this sample at fixed cosmology ($\Omega_m=0.31$, $h=0.68$, $\sigma_8=0.77$).  Again, the large scales we probe are dominated by the linear contributions.

Unfortunately the Planck lensing maps are relatively noisy, leading to broad posteriors on many parameters.  The grid of models used to train \texttt{Anzu} is not wide enough to encompass the full posterior, so we cannot use \texttt{Anzu} in its present form for our cosmological inferences.  However, for models within the training volume we expect \texttt{Anzu} to be more accurate than CLEFT and so we have employed \texttt{Anzu} throughout the analysis as a systematic check.

\subsection{Parameter priors, sampling and emulation}
\label{sec:priors}

For our parameter inference we employ a now-standard Markov Chain Monte Carlo technique, using the \texttt{Cobaya} code \cite{Torrado21,CobayaSoftware} with the \texttt{CosmoMC} sampler \cite{Lewis02,Neal05,Lewis13} and \texttt{GetDist} \cite{Lewis19} visualization software.  For each point in parameter space we generate $C_\ell^{XY}$ using the Limber approximation from the underlying real-space power spectra and use a Gaussian likelihood.

The evaluation of the theoretical models involves computation of the linear theory power spectrum (for which we use CAMB \cite{Lewis00,Howlett12}) and higher order corrections as described above.  The codes for doing this are quite efficient, but still take several seconds per call.  When many chains need to be run for many samples with many different settings per chain this can add up to a large computational cost.  In order to make the chains more efficient we do not call the theory model directly, but rather use an additional emulation/interpolation stage based on neural networks.  Neural networks can be used as ``universal'' function interpolators, whose accuracy is limited primarily by the amount of data available to train them.  In this regard the combination of neural networks and perturbative models is particularly attractive, because it is possible to generate millions of models to train the network at modest computational cost.  In a sense we pay the cost of evaluating models in a given family up front, and such models can be used within many chains.  This technique is gaining currency in cosmology, and similar approaches have been presented in refs.~\cite{Auld08,Agarwal14,Albers19,Manrique20,Kasim21,Arico21,Angulo21,Mancini21,Veronesi21}.

We present the details of the neural network architecture in ref.~\cite{Emulasaur}.  Briefly we use an architecture and activation function similar to that in ref.~\cite{Alsing20}.  The network has $4$ fully connected layers of $128$ neurons, with an activation function \cite{Alsing20}
\begin{equation}
    a(\mathbf{x}) = \left[ \mathbf{\gamma} + \frac{1-\mathbf{\gamma}}{1+\exp(-\mathbf{\beta}\odot\mathbf{x})} \right] \odot \mathbf{x}
\end{equation}
where $\odot$ indicates element-wise multiplication, $\beta$ and $\gamma$ are parameters and $\mathbf{x}$ are the inputs.  A final layer performs a linear offset and scaling.  A grid of $10^6$ models covering a broad range of cosmological and bias parameters and redshifts is generated.  The spectra are normalized, the mean subtracted and the arcsinh taken to reduce the dynamic range, and then compressed into $64$ principal components.  The coefficients of these principal components are then predicted by our neural network.  We find the neural network is able to reproduce the spectra in our training and test sets with a median error of $0.1\%$ with the prediction step now several orders of magnitude faster than the theory codes.  We find the posterior contours using the full model and the emulator are indistinguishable in all tests we have performed.

\begin{table}[]
    \centering
    \begin{tabular}{cc|cc}
    \multicolumn{2}{c|}{$\quad$Cosmological$\quad$} &
    \multicolumn{2}{c}{Nuisance} \\
    \hline
    $\theta_\star$& 0.0104109
    & $b_1$       & $\mathcal{U}(0,3)$ \\
    $n_s$         & 0.97
    & $b_2$       & $\mathcal{U}(-5,5)$ \\
    $\tau$        & 0.07
    & $\alpha_a$  & $\mathcal{N}(0,50)$ \\
    $\omega_b$    & 0.022
    & $\alpha_\times$  & $\mathcal{N}(0,50)$ \\
    $\omega_c$    & $\mathcal{U}(0.08,0.16)$
    & SN          & $\mathcal{N}({\rm Poiss},30\%)$ \\
    $\ln(10^{10}A_s)$ & $\mathcal{U}(2,4)$
    & $s_\mu$     & $\mathcal{N}(-,0.1)$ \\
    \hline
    \end{tabular}
    \caption{Priors used in our fits.  The first 2 columns contain the priors on cosmological parameters, with the symbols having their usual meanings (e.g.\ ref.\ \cite{PlanckParams18}).  The last 2 columns show the nuisance parameters (biases and counterterms) for the PT model described in \S\ref{sec:pk_models}.  The bias terms $b_1$, $b_2$ and the slope $s_\mu$ are dimensionless, the counterterms ($\alpha_i$) are in $h^{-2}\mathrm{Mpc}^2$ and the shotnoise is in $h^{-3}\mathrm{Mpc}^3$.  A numerical value indicates a fixed parameter, $\mathcal{U}$ is the uniform distribution while $\mathcal{N}$ is a normal distribution.  For the shotnoise (SN) we use the shorthand ``Poiss'' to refer to the Poisson value and for $s_\mu$ we use $-$ to refer to the mean value, which are given in Table \ref{tab:samples} for each sample.
    }
    \label{tab:priors}
\end{table}

Especially for the PT model, for which we will have limited dynamic range, some of the higher order bias parameters are degenerate with the stochastic term (i.e.\ the shot-noise).  For this reason we apply loose priors to these parameters to improve the efficiency of the chains (Table \ref{tab:priors}).  We can gain some intuition about the expected size of our bias parameters by considering Fig.\ 8 of ref.\ \cite{Abidi18}, generalized to mock galaxies in refs.\ \cite{Barreira21,Zennaro21}.
For the shot-noise term we choose a Gaussian prior on SN, centered on the Poisson value (Table \ref{tab:samples}) with a width of $30\%$ \cite{Kokron21b}.
For the slope of the number counts ($s_\mu$) we use a Gaussian prior centered on the measured value (Table \ref{tab:samples}) with a width of $0.1$.  This is approximately an order of magnitude larger than the statistical uncertainty in the determination of $s_\mu$ \cite{PaperI}.
For the cosmological parameters and several of the nuisance parameters the priors are broad enough to be irrelevant, however in the case of the counterterms the posterior depends upon the imposed shot-noise prior and our constraint on $s_\mu$ is determined entirely by the prior.  With only one exception (see \S\ref{sec:results}) we find these nuisance parameters do not influence the cosmological ones, so our cosmological constraints are very weakly dependent on our prior choices for these parameters.  We hold the primordial power spectrum slope, $n_s$, and the optical depth, $\tau$, fixed.  We are quite insensitive to the specific choices: changing these by $2\,\sigma$ does not appreciably alter our constraints.

The fact that our data are projected, and thus do not strongly resolve the baryon acoustic oscillations that can be used as a standard ruler, and of limited dynamic range means that we are sensitive primarily to a combination of $\Omega_m$ and $\sigma_8$ rather than each individually.  When doing fits to $\Lambda$CDM parameters we hold the angular size of the sound horizon, $\theta_\star$, fixed.  This parameter is known to exquisite precision and depends very weakly on model assumptions as it comes from the observed locations of the CMB acoustic peaks \cite{PlanckParams18} (changes in the precise value of $\theta_\star$ hardly impact our constraints, nor does holding $H_0$ fixed instead of $\theta_\star$).  We additionally jointly fit to the Pantheon SNe data compilation \cite{Scolnic18} and BAO data \cite{SixDF,SDSS_DR7,BOSS_DR12}.  The combination of the SNe data and $\theta_\star$ provides relatively strong constraints on $\Omega_m$, constraining it to be $\Omega_m\approx 0.30\pm 0.02$ in agreement with a host of other observations \cite{Zyla:2020zbs}.  The combination with BAO approximately halves the error as the BAO data provide a consistent, but tighter constraint on $\Omega_m$, while being primarily a geometric measurement.  This better constrains the parameter combination that our auto- and cross-correlations do not constrain well themselves and improves the convergence of the chains by downweighting low-$\Omega_m$ solutions that are allowed by just our data.  When fitting the $\Lambda$CDM model to the combination of the 4 redshift slices we have 52 data points and 26 parameters, 9 of which are at least partially prior dominated.

\section{Simulations and pipeline test}
\label{sec:simulations}

We use mock catalogs \cite{MockPaper} derived from N-body simulations in order to validate our models and determine our scale cuts.  The theoretical models have been extensively compared to simulations and data elsewhere \cite{Modi17,Chen20,Chen21a,Kitanidis21,Krolewski21,Kokron21a} but we include these tests to validate the full pipeline for samples similar to our data.  Since none of our data, covariance or models are calibrated from these simulations our requirements are relatively relaxed; we desire a sufficiently complex input model with approximately the same redshift distribution and clustering power.

\subsection{Mock catalogs}

The mock samples are generated from the \texttt{Buzzard} N-body simulations \cite{DeRose2019, DeRose2021, Wechsler2021} using the \texttt{Addgals} algorithm, which assigns galaxy positions, velocities, rest-frame absolute magnitudes and spectral energy distributions to each galaxy. This allows us to generate photometry in the DECAM bandpasses, so that we can make galaxy sample selections similar to those used in our measurements on the data. In particular, we make small adjustments to the color cuts used to select the DESI LRG sample in order to match the angular number density of the LRG sample. After tuning to the total angular number density, our selection in the simulations agrees well with the comoving number density and projected clustering, $w_p$, for the LRGs measured during survey validation \cite{MockPaper}. In order to place our simulated sample into redshift bins, we assign photometric redshift estimates to each galaxy. We do this by first binning the DESI LRG data into narrow ($\Delta z=0.075$) redshift bins, and estimating the mean photometric redshift uncertainty in these narrow bins, providing a model for $\sigma_{z}(z)$. For each simulated galaxy, we then assign a photometric redshift by perturbing the true redshift by a Gaussian error with standard deviation given by $\sigma_{z}(z)$. We use the Born approximation and the dark matter distribution in the simulation to produce mock CMB-lensing $\kappa$ maps.  The angular number densities of the mock samples are 85, 155, 164, and $145\,{\rm deg}^{-2}$, very close to those of the data (Table \ref{tab:samples}).

\subsection{Mock spectra}

\begin{figure}
    \centering
    \resizebox{\columnwidth}{!}{\includegraphics{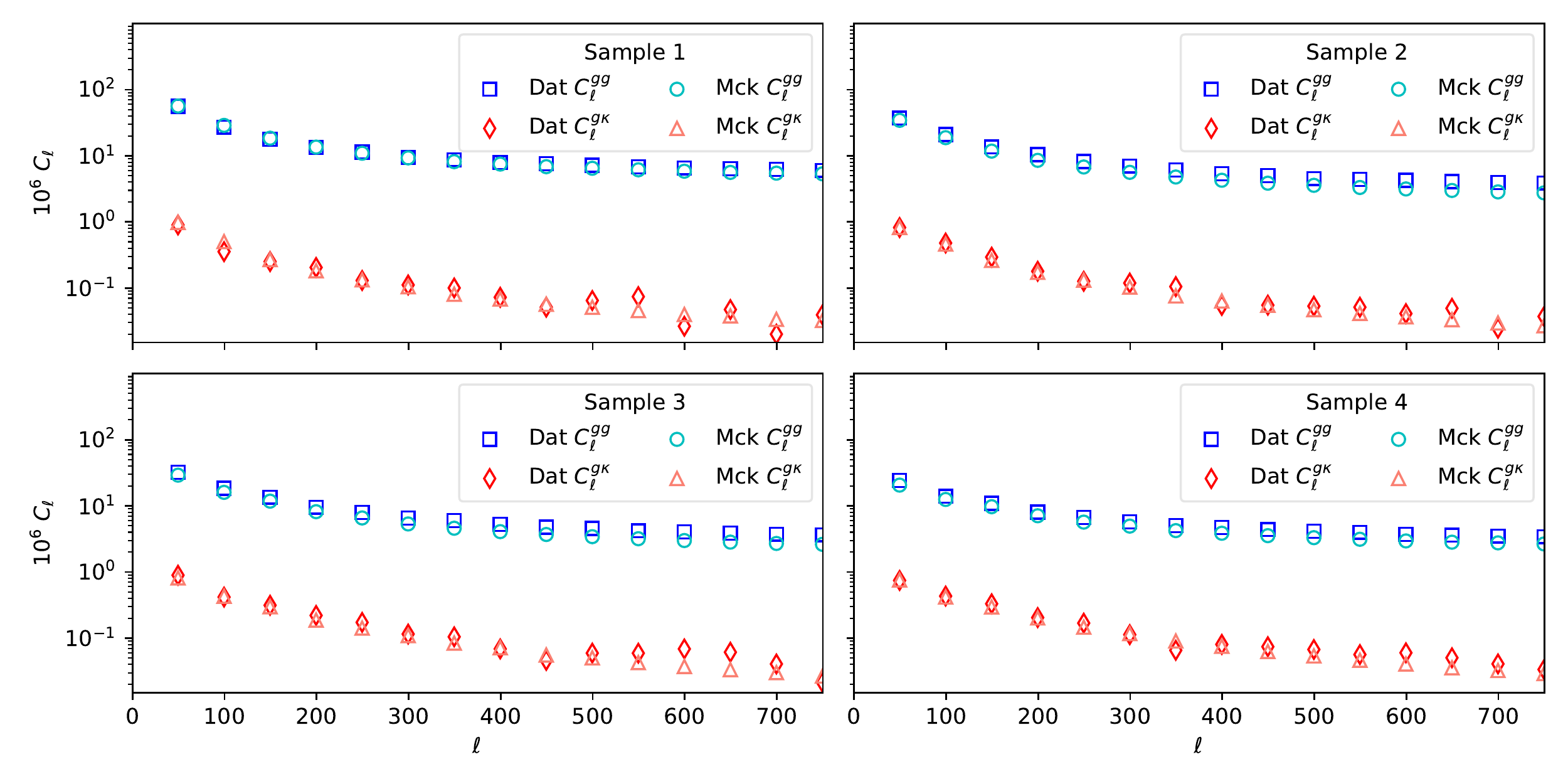}}
    \caption{A comparison of the angular power spectra of the data and the mocks that we use to validate our pipeline.  For each of the 4 samples we show $C_\ell^{gg}$ and $C_\ell^{g\kappa}$ for both the mocks and data, and we see that the agreement is very good.  We have omitted the error bars to avoid crowding, but they are the same as in Fig.~\ref{fig:bandpowers}.  Though we do not show it here, the number density and $dN/dz$ of the mocks and data are also in good agreement.}
    \label{fig:mock_comparison}
\end{figure}

To improve their value as a cross-check we ensured that the pipeline to create the mock measurements was completely independent of the fitting pipeline, though we computed the auto- and cross-spectra using \texttt{NaMaster} (with the same binning as for the data) in both cases.  A comparison of $C_\ell$ between data and mock is shown in Fig.~\ref{fig:mock_comparison}. For all simulated analyses we use the the mean of the measurements made from 7 quarter-sky \texttt{Buzzard} simulations.

\begin{figure}
    \centering
    \resizebox{0.48\columnwidth}{!}{\includegraphics{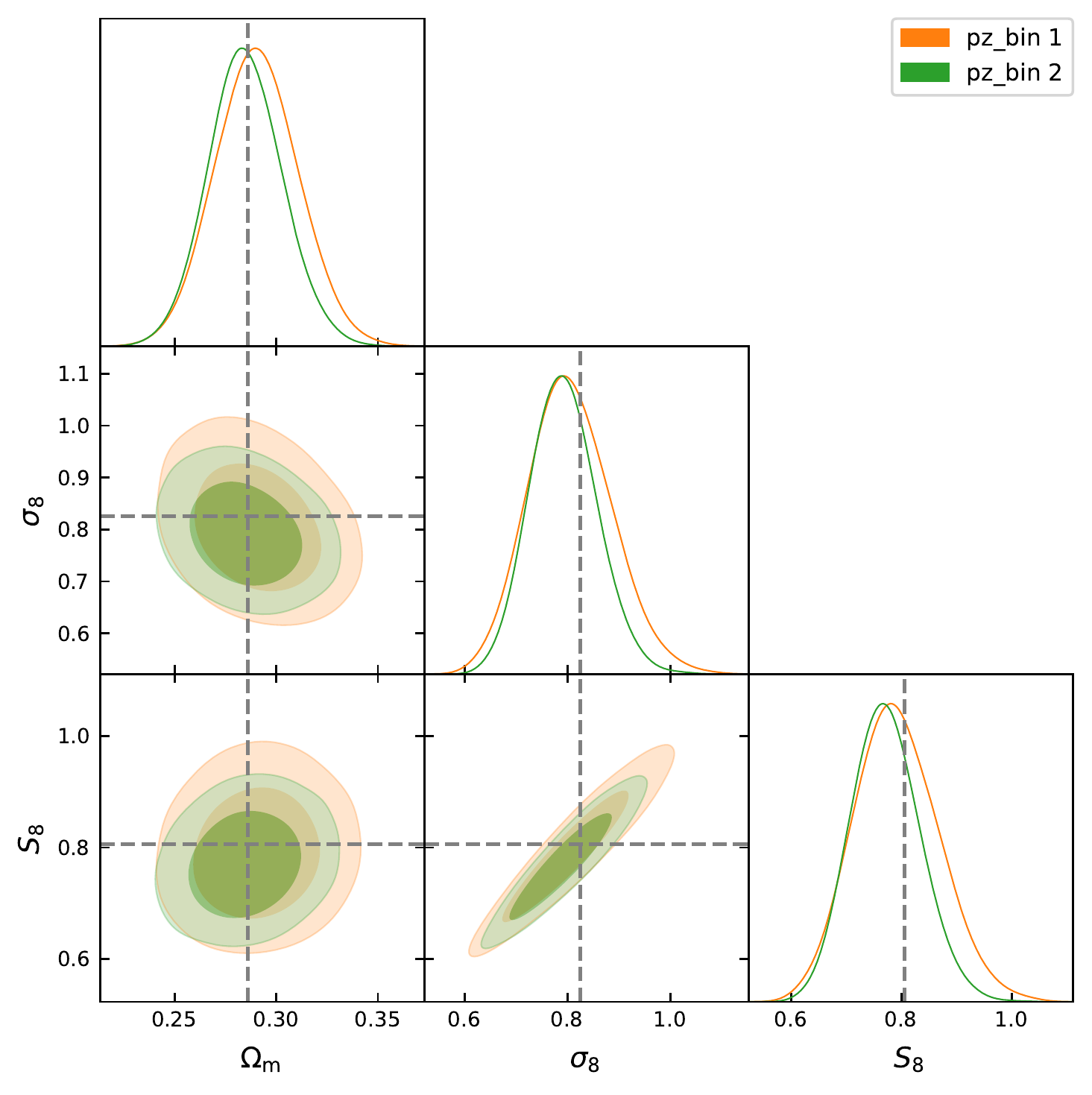}}
    \resizebox{0.48\columnwidth}{!}{\includegraphics{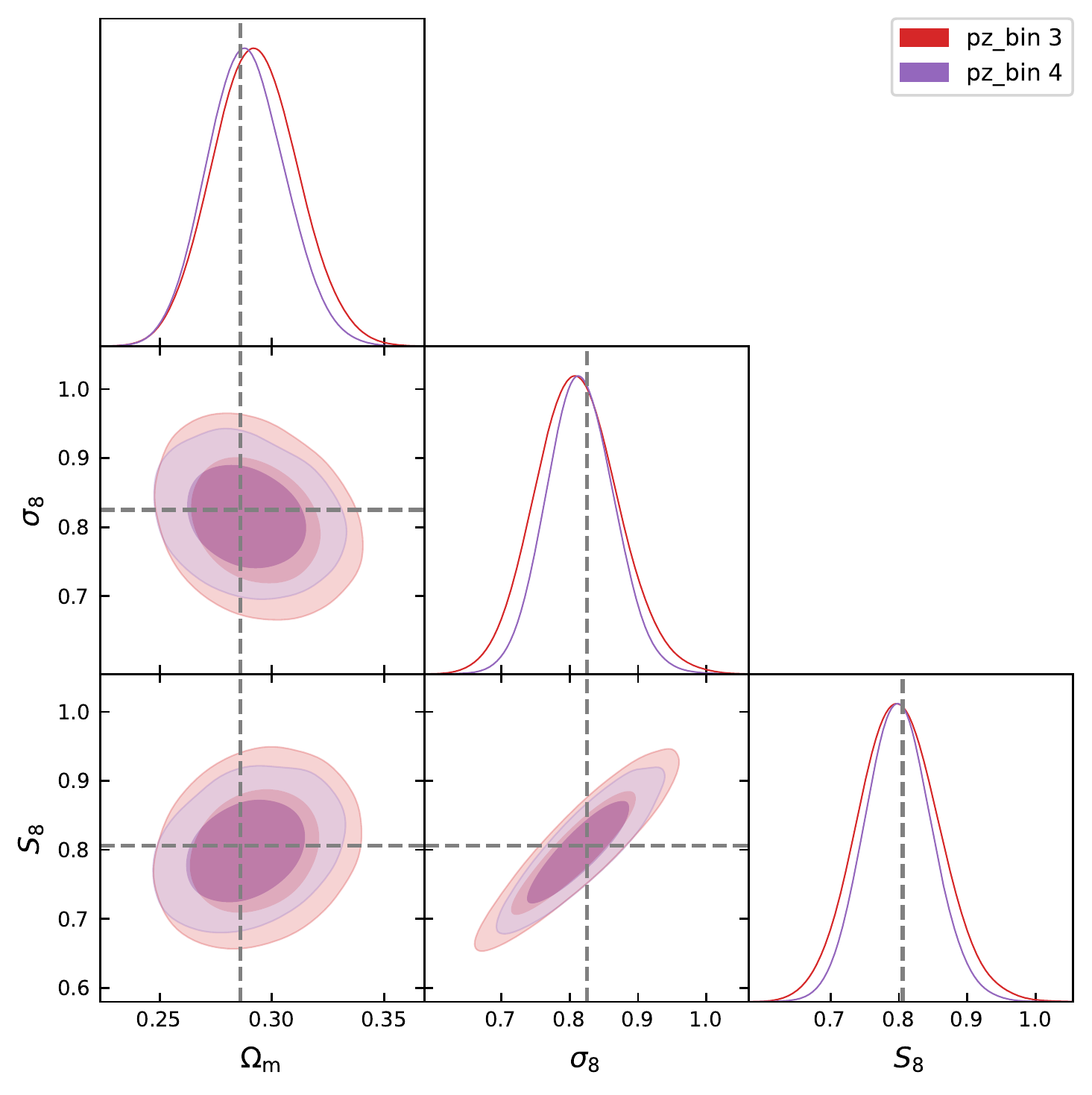}}
    \caption{Fits of our $\Lambda$CDM models, with fiducial scale cuts as in Table \ref{tab:samples}, to the mocks.  The fits are nearly unbiased in all of the parameters, with the differences being consistent with residual noise in the mock data arising from finite volume (see text). }
    \label{fig:mock_validation}
\end{figure}

We then fit the mock data with our models.  For this test we used the window functions and data from the mocks and the covariance matrix of the data.  Since the \texttt{Buzzard} simulations are pure dark matter simulations we set $m_\nu=0$ and similarly match $n_s$, $\Omega_b\,h^2$ and the central value of the shot-noise to the values used in the simulations.  To mimic the constraints from the SNe we generate a mock dataset with the same magnitude errors and redshifts as the Pantheon data but distances calculated using the mock cosmology.  Our pipeline returns unbiased estimates of the cosmological parameters for all 4 of our mock samples, as shown in Fig.~\ref{fig:mock_validation}.  Since the volume of the mock catalogs is approximately $3.5\times$ the volume of the data we would expect biases in the parameters at $\approx\frac{1}{2}\sigma$, and so regard agreement at this level as a successful cross-check of the model.  The pipeline is also unbiased for the combination of the 4 samples (not shown in the figure to avoid crowding) with $\Omega_\mathrm{m} = 0.285\pm 0.014$ and $\sigma_8 = 0.800\pm 0.037$ to be compared to $0.286$ and $0.825$.  The best-fit model has $\chi^2\simeq 6.6$ for the clustering data (for 52 data points and 26 parameters, 9 of which are at least partially prior dominated) on a volume $3.5\times$ larger than our survey.  Note that our data do not strongly constrain $\Omega_m$ and $\sigma_8$ individually, but are primarily sensitive to a combination as mentioned above.  Movement along the $\Omega_m-\sigma_8$ degeneracy direction incurs only a small $\chi^2$ penalty.

\subsection{Blinding}

Our analysis was performed blind to the cosmological parameters.  We did not blind the input catalogs, weights or maps nor did we blind the angular power spectra or covariances (though the pipeline is straightforward and there are no user-defined ``knobs'' aside from binning).  The choice of samples, maps and masks were fixed before any cosmology fits were done. Similarly, our choice of priors and range of scales to be fit for each model was determined in advance of fitting the data using measurements from mock catalogs (described above).  On the few occassions when tests of the full pipeline were performed on the data (primarily to test the code for bugs in I/O), we ``blinded'' the $dN/dz$ used in the inference by applying an unknown affine transformation to $z$ before we did any fits and removed tick labels from any plots.  Once the data were unblinded we made no changes to these inputs or choices.

We flagged the \texttt{pz\_bin} 1 sample as being sensitive to our choice of foreground subtraction in the CMB $\kappa$ map before unblinding, but did not find any reasons to remove it at that time.  After unblinding it was found to lead to noticeably lower $\sigma_8$ than the other samples, though it is not highly discrepant statistically given its large error bar (see below).  We have chosen to include this sample in our fiducial results though future work should investigate this sample more closely.

\section{Results}
\label{sec:results}


We now describe the fits to the galaxy autospectra and the galaxy-convergence cross-spectra for the 4 samples individually and combined.  We begin by showing fits to the amplitude at fixed power spectrum shape, before discussing fits to the $\Lambda$CDM model.

\subsection{Fixed shape}

Within the $\Lambda$CDM paradigm, CMB anisotropies largely fix the epoch of matter radiation equality, the slope of the primordial power spectrum and the physical densities of baryons, cold dark matter and neutrinos \cite{PlanckParams18}.  These are the parameters that determine the shape of the power spectrum \cite{PlanckLegacy18}.  Between $z\approx 10^3$ and today the linear theory power spectrum shape changes only modestly (due to neutrinos going non-relativistic) while the amplitude grows by orders of magnitude.  Testing whether the growth predicted by the $\Lambda$CDM model matches current observations is one of the goals of this paper.

\begin{figure}
    \centering
    \resizebox{0.48\columnwidth}{!}{\includegraphics{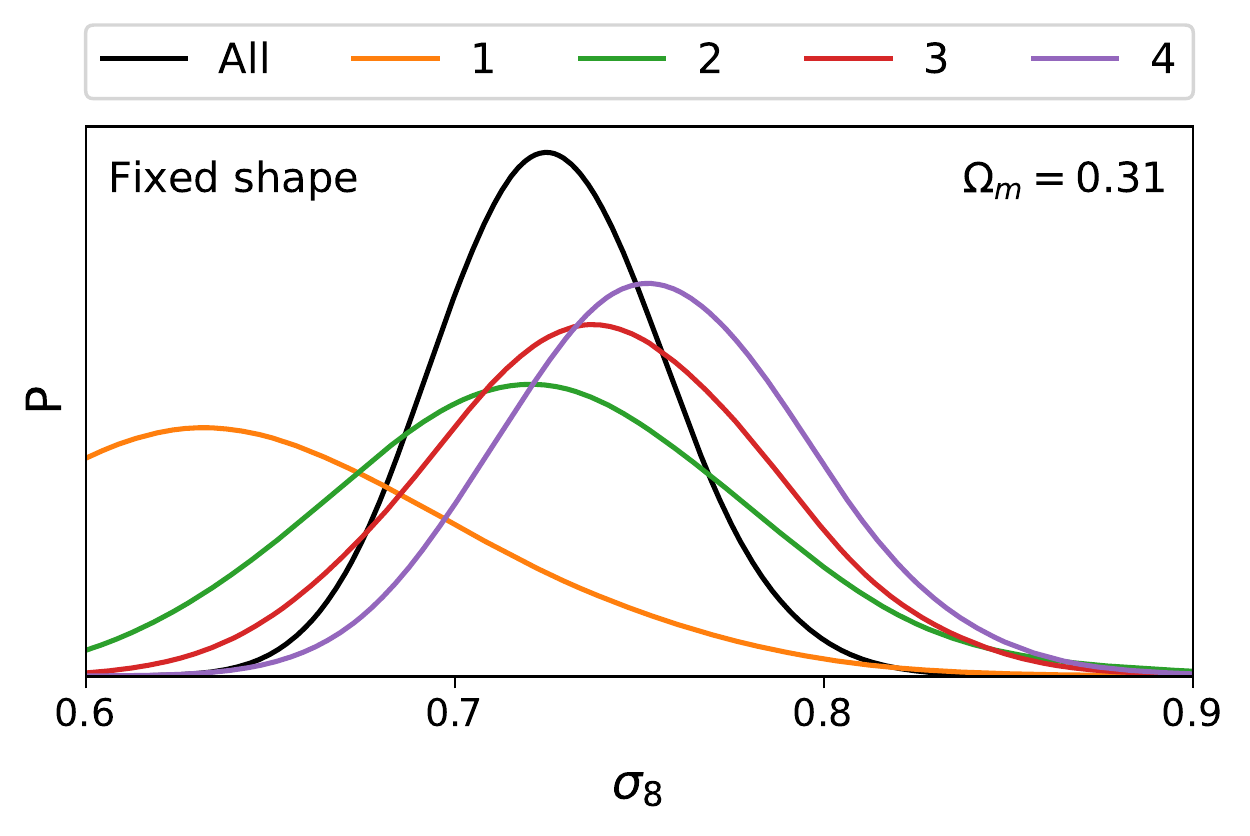}}
    \resizebox{0.48\columnwidth}{!}{\includegraphics{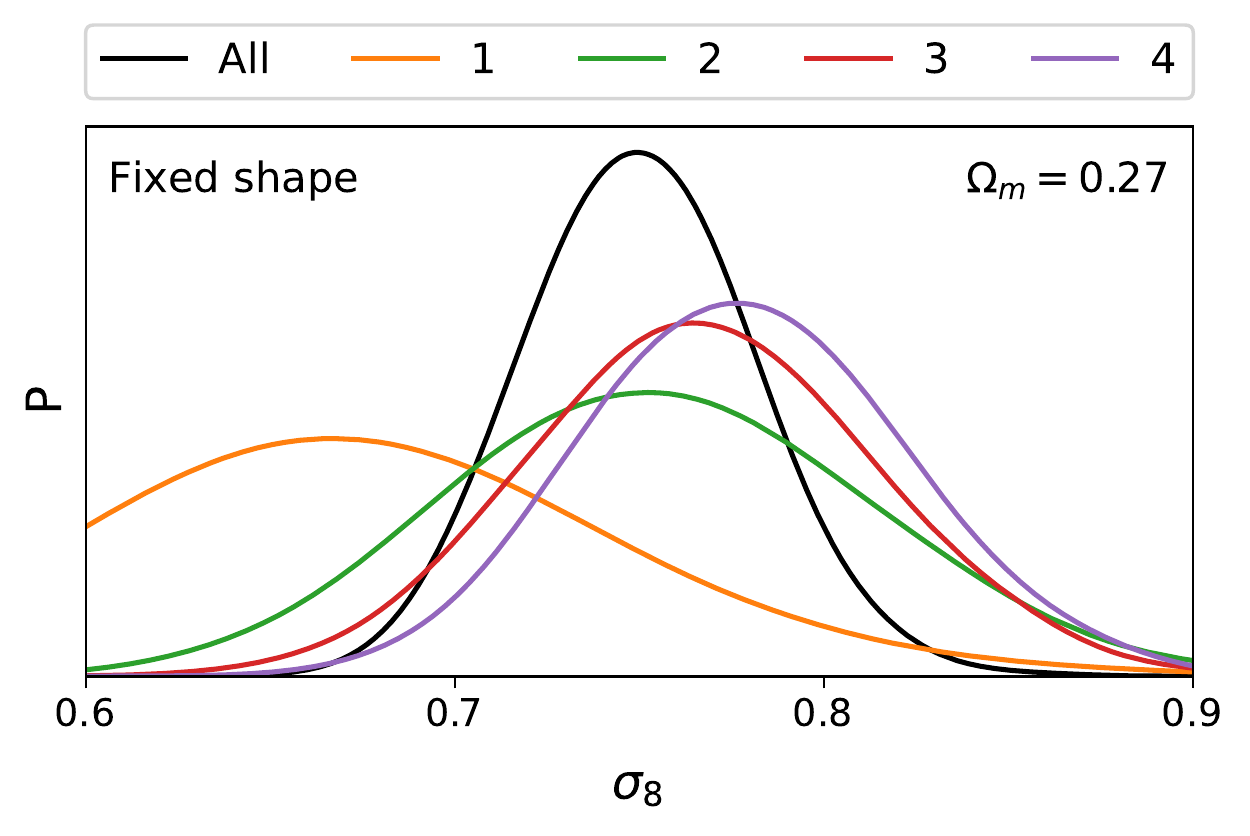}}
    \caption{Posteriors for the power spectrum amplitude ($\sigma_8$) in the ``fixed shape'' fits.  We present the results for two assumptions about the matter density, $\Omega_m=0.31$ (left) and $\Omega_m=0.27$ (right), as described in the text.}
\label{fig:fixed_shape}
\end{figure}

Our data, like other lensing surveys, are consistent with the power spectrum shape determined from the CMB but unfortunately do not provide competitive (compared to Planck \cite{PlanckParams18}) constraints on this shape.  Our spectra best constrain a combination of the matter density and the clustering amplitude.  One consistency test of the model is therefore to hold the background cosmology and power spectrum shape fixed while allowing the low $z$ amplitude determined by lensing to vary and comparing the results to that predicted by $\Lambda$CDM fit to the CMB data.  We do this for two values of the matter density, differing by about 10\%.

For the chain with $\Omega_m\simeq 0.31$ we find $\sigma_8=0.640 \pm 0.066$, $0.723 \pm 0.057$, $0.739 \pm 0.047$ and $0.755 \pm 0.042$ for bins 1-4.  The combined constraint (including the inter-sample covariance) is $0.73 \pm 0.03$.  Note there is a tendency for the higher redshift bins to prefer higher $\sigma_8$, i.e.\ the observed level of clustering is not growing as rapidly as the $\Lambda$CDM model predicts.  For the $\Omega_m\simeq 0.27$ chain the corresponding values are $0.675 \pm 0.072$, $0.756 \pm 0.059$, $0.767 \pm 0.048$ and $0.779 \pm 0.044$ for bins 1-4 and $0.75 \pm 0.03$ for the full sample.  The clustering amplitude is higher due to the lower matter density, and the trend of increasing best-fit $\sigma_8$ with redshift persists.  The marginalized posterior distributions for the 4 bins and the combined sample are shown in Figure \ref{fig:fixed_shape} for each of the two values of $\Omega_m$.  We find that our data constrain a combination of $\Omega_m$ and $\sigma_8$ that is intermediate between the $\Sigma_8=\sigma_8(\Omega_m/0.3)^{0.25}$ constrained by CMB lensing autocorrelations and the $S_8=\sigma_8(\Omega_m/0.3)^{0.5}$ constrained by recent cosmic shear surveys.  To avoid introducing yet another summary statistic, we shall quote $\Sigma_8$ and $S_8$ constraints below when comparing to CMB and cosmic shear results.

The very low $\sigma_8$ preferred by the lowest redshift slice prompts further investigation.  This is also the point with the lowest S/N and smallest dynamic range in scale.  To assess goodness-of-fit and check for ``volume effects'' we can also look at the profile likelihood.  As this lowest redshift slice has a very small contribution from magnification and is sample variance dominated, our constraints on the shot-noise and magnitude slope are prior dominated.  Holding those parameters fixed at the centers of the prior and holding the cosmology fixed we can find the minimum $\chi^2$, varying the bias and counterterms.  There are 10 data points and 4 free parameters, for 6 degrees of freedom (or between 5 and 6 if we account for the inflation of the error bar in the autospectrum at $\ell=50$ due to beyond-Limber contributions).  By varying the values of $\sigma_8$ in the `fixed' cosmology for each run we can compute $\chi^2_{\rm min}(\sigma_8)$ and hence $\mathcal{L}_{\rm profile}(\sigma_8)$.  Taking $\Omega_m\simeq 0.31$ we find for the Planck-preferred $\sigma_8\simeq 0.81$ that $\chi^2_{\rm min}=9.3$ (PTE: 16\%).  This becomes $7.1$ (31\%) for $\sigma_8=0.75$, $6.1$ (42\%) for $\sigma_8=0.70$ and $5.8$ (45\%) for $\sigma_8=0.65$.  While the $\Delta\chi^2$ is significant, all of these values are statistically acceptable, indicating that for the scales analyzed this bin doesn't provide much constraining power.  Indeed the lack of dynamic range in scale is partly responsible for the low value of $\sigma_8$ we obtain for this bin, as we find that $\sigma_8$ is quite degenerate with $\alpha_\times$ with low $\sigma_8$ corresponding to large $\alpha_\times$.  With $\alpha_\times$ largely unconstrained by the low $\ell$ data, the marginal posterior turns out to be weighted to low $\sigma_8$ by the integration over $\alpha_\times$.  Our prior is broad enough that even large corrections from $\alpha_\times$ are permitted statistically, indicative of a breakdown of perturbation theory for the lowest $\sigma_8$ values.  Unfortunately the data are not constraining enough to keep $\alpha_\times$ in the consistent region without a strong prior; fortunately the low constraining power of this redshift bin means our final results are not much affected by this issue.

\subsection{CDM}

\begin{figure}
    \centering
    \resizebox{\columnwidth}{!}{\includegraphics{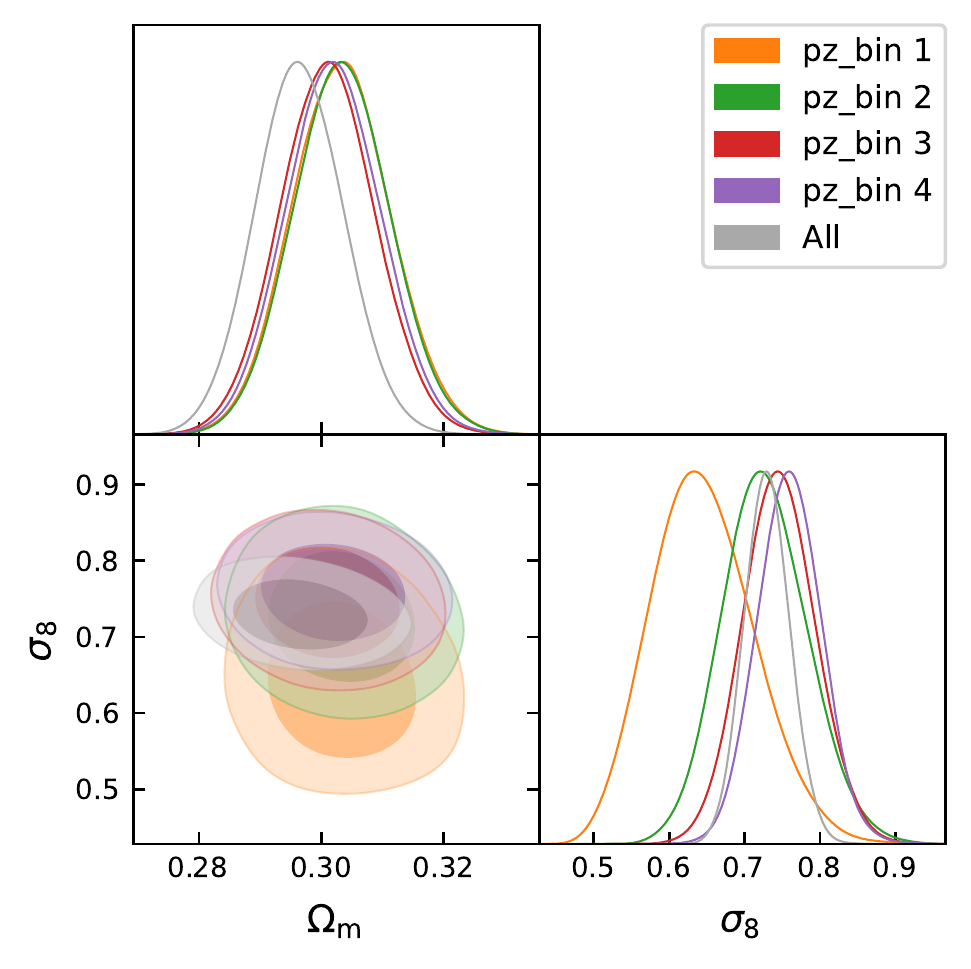}}
    \caption{Constraints on key $\Lambda$CDM parameters from each of our 4 \texttt{pz\_bin} samples, and from the combination.  Our constraint is better described as a measurement of $\Sigma_8=\sigma_8(\Omega_m/0.3)^{0.25}$, as in CMB lensing, than as a constraint on the $S_8$ more often constrained by low redshift cosmic shear experiments. }
    \label{fig:data_corner}
\end{figure}

\begin{figure}
    \centering
    \resizebox{\columnwidth}{!}{\includegraphics{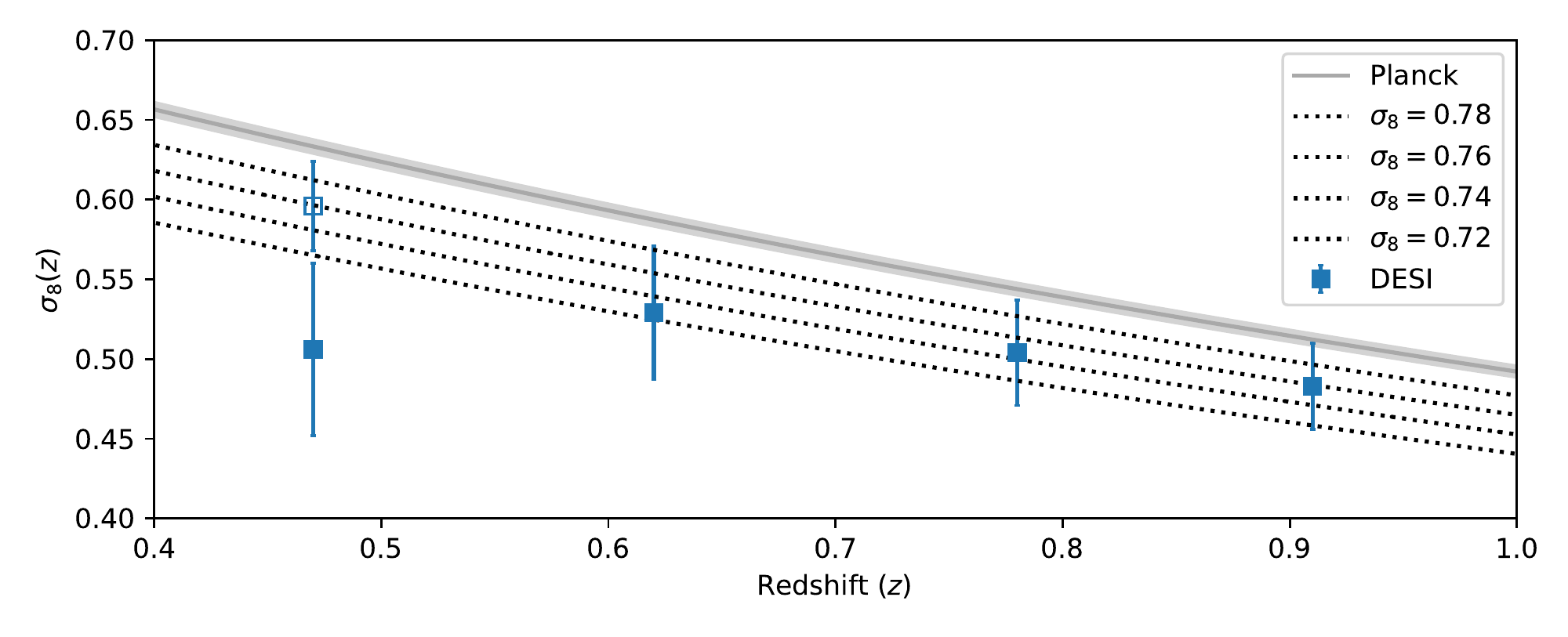}}
    \caption{Constraints on $\sigma_8(z)$ from the DESI data and from Planck.  In each case the uncertainty in the $\Lambda$CDM model parameters is propagated into an uncertainty in the prediction for $\sigma_8(z)$ as a function of redshift.  For Planck the dark grey line indicates the mean prediction while the light grey band covers the $\pm 1\,\sigma$ uncertainty from the combination Planck primary CMB anisotropies and CMB lensing.  For the DESI data the constraints are from each \texttt{pz\_bin}, plotted at the value of $z_{\rm eff}$ for each sample.  The open point at $z\simeq 0.47$ corresponds to the constraint with $\alpha_\times=0$, which eliminates the projection effect described in the text.  The DESI data fall short of the Planck predictions at all redshifts, with a weak preference for slower growth than $\Lambda$CDM predicts.  The dotted lines show the $\Lambda$CDM prediction for $\Omega_m=0.3$ and $\sigma_8=0.72$, 0.74, 0.76 and 0.78 to guide the eye. }
    \label{fig:sigmaz}
\end{figure}

In order to asses the extent to which uncertainties in the matter density and shape of the power spectrum affect our amplitude constraints, we also fit our data directly to the predictions of the $\Lambda$CDM model.  For these fits we vary the matter density as well as the power spectrum amplitude, using the known angular scale of the CMB ($\theta_\star$) to fix $h$ as a function of $\Omega_m$.  Although we cover a substantial range in redshift, so we have some contour rotation to break degeneracies, our data still do not strongly constrain movement along the $\Omega_m-\sigma_8$ degeneracy (see appendix \ref{app:noext}).  Fortunately, several other data sets do provide tight constraints on this dimension within the $\Lambda$CDM framework.  We combine our data with the Pantheon SNe data sample and a combination of BAO data to better constrain $\Omega_m$.  Figure \ref{fig:data_corner} shows the constraints on the $\Lambda$CDM parameters from each of the 4 samples individually, as well as the combined constraint (including the intersample covariance).

Our data can be most conveniently summarized as a constraint on $\Sigma_8=\sigma_8(\Omega_m/0.3)^{0.25}$, with bins 1-4 giving $0.647 \pm 0.068$, $0.729 \pm 0.057$, $0.747 \pm 0.048$ and $0.761 \pm 0.042$ or as a constraint on $S_8=\sigma_8(\Omega_m/0.3)^{0.5}$ with bins 1-4 giving $0.649 \pm 0.068$, $0.731 \pm 0.057$, $0.747 \pm 0.048$ and $0.762 \pm 0.043$ respectively.  Combining all 4 samples we find $\Sigma_8=0.727 \pm 0.030$ or $S_8=0.725 \pm 0.030$.  The best-fit model has $\chi^2\simeq 23.5$ for the clustering data (for 52 data points and 26 parameters, 9 of which are at least partially prior dominated).  While not strongly inconsistent with the other three, the first bin is clearly lower than the other three which are quite compatible with each other.  As described above this is partly a ``volume effect''.  The lowest bin, having the smallest dynamic range in scale, cannot strongly constrain $\alpha_\times$.  This parameter is in turn correlated with $\sigma_8$.  Setting $\alpha_\times=0$ gives a reasonable fit to the data with $\Sigma_8=0.76\pm 0.04$ indicating that the low value for the full model is partly a result of parameter volume effects in the posterior.  If we omit this first redshift bin, our constraint becomes $\Sigma_8=0.742 \pm 0.032$.

We noted above the tendency of the best-fit $\sigma_8$ to increase with increasing sample redshift.  This is responsible for the shift to lower $\Omega_m$ of the combined samples constraint compared to each of the individual samples (Fig.~\ref{fig:data_corner}).  Another view of this is shown in Figure \ref{fig:sigmaz} which plots constraints on $\sigma_8(z)$, marginalized over the other parameters, as a function of redshift.  The grey lines show the predictions of the $\Lambda$CDM model conditioned on the Planck primary CMB plus lensing data (with the dark grey line being the mean and the grey band encompassing the $\pm 1\,\sigma$ uncertainty).  All of the points lie below the line, with the lowest redshift points deviating the most strongly.  We also show (dotted lines) the $\Lambda$CDM prediction for $\Omega_m=0.3$ and $\sigma_8=0.72$, 0.74, 0.76 and 0.78 for reference.  While of relatively low statistical significance, the data suggest structure that is not growing as rapidly as $\Lambda$CDM predicts with the largest shortfall appearing below $z\approx 0.6$.

Finally, while we are unable to use \texttt{Anzu} for the lowest redshift slices (where it would be the most valuable) due to the likelihood having non-trivial support outside of the range of validity of the emulator, we can compare the results from \texttt{Anzu} to our LPT model for the higher redshift points.  Fitting to the same $\ell_{\rm max}$ the agreement on parameters is excellent.  For example, for \texttt{pz\_bin} 4 the constraints differ by less than one percent between the two models, significantly smaller than our statistical errors.  Given the kernels in Fig.~\ref{fig:kernel} this is reassuring.  If we increase the fitting range from $\ell_{\rm max}=400$ to $\ell_{\rm max}=600$ the constraint on $\sigma_8$ also changes by less than 1\%, and the error shrinks negligibly.  This is not unexpected, the Planck lensing map is signal dominated only at very low $\ell$ and the galaxies become shot noise limited for $\ell>425$.  We would expect more gain from \texttt{Anzu} in the lower redshift bins, where the non-linear scale is shifted to much smaller $\ell$ (Fig.~\ref{fig:kernel}).  Unfortunately, for these bins some of the support for the likelihood lies outside of the grid of simulations upon which the \texttt{Anzu} emulator was trained, so the emulator error is not well controlled.  We aim to increase the coverage of \texttt{Anzu} with more simulations in the future.

\subsection{Comparison}
\label{sec:comparison}

\begin{figure}
    \centering
    \resizebox{\columnwidth}{!}{\includegraphics{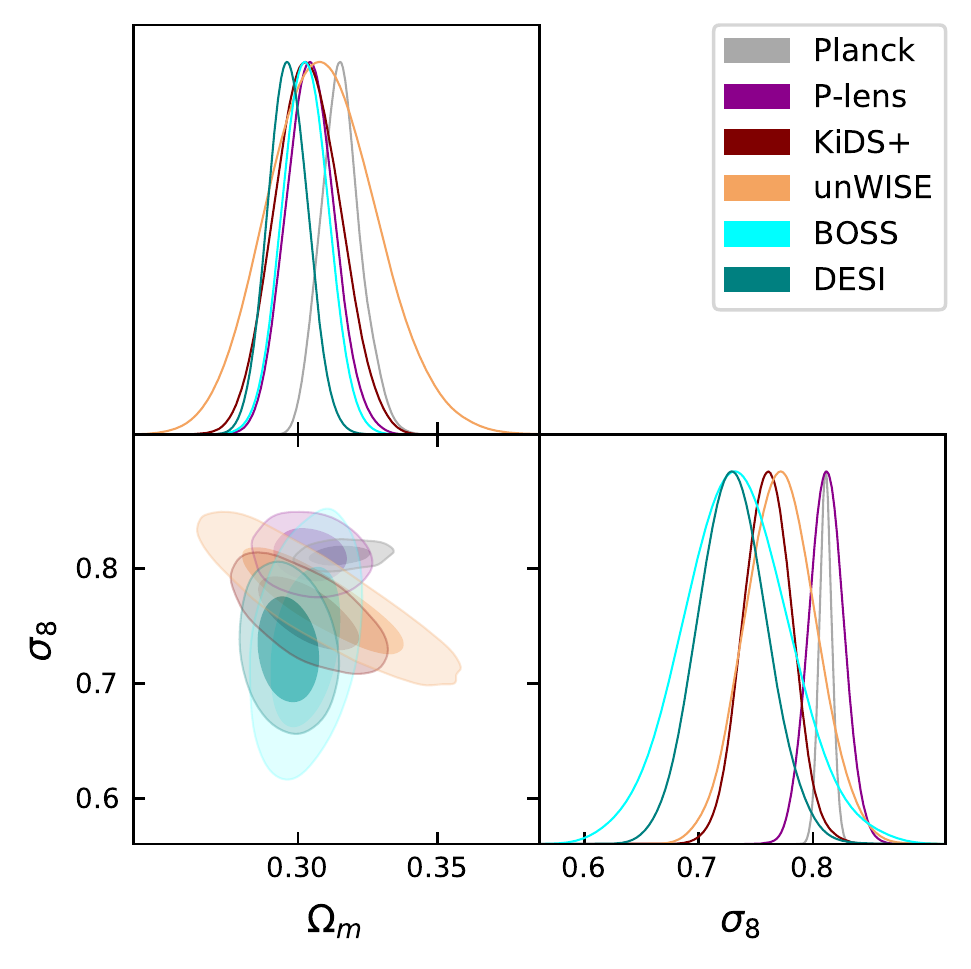}}
    \caption{Posteriors for two key $\Lambda$CDM parameters constrained by our data compared to a variety of recent experiments (as shown in the legend and described in the text).  Though the models fitted, priors assumed and methodologies differ the comparison paints a picture of the level of agreement of different low redshift probes of the growth of large-scale structure within the context of $\Lambda$CDM and their comparison to primary CMB anisotropies (in this case, Planck).  See text for further discussion. }
    \label{fig:cmp_corner}
\end{figure}

Figure \ref{fig:cmp_corner} shows that our result gives a ``low'' clustering amplitude, compared to Planck.  In terms of $\Sigma_8\equiv\sigma_8(\Omega_m/0.3)^{0.25}$ we find $\Sigma_8=0.727\pm 0.030$ for our full sample.  Including CMB lensing, Planck finds $\Sigma_8 = 0.821\pm 0.008$ \cite{PlanckParams18}.  This corresponds to a $3\,\sigma$ ``tension'' with Planck when adding the statistical uncertainties in quadrature.  We have not included the Planck lensing auto-correlation in our fits.  Once BAO and weak priors are included that auto-correlation (P-lens in figure \ref{fig:cmp_corner}) gives $\Sigma_8 = 0.815\pm 0.016$ \cite{PlanckLens18}.  Our result is $2.5\,\sigma$ lower than this, with approximately twice the uncertainty.  If we had included the $\kappa$ autospectrum (more closely following the ``$3\times 2$pt'' strategy often employed by cosmic shear surveys) our result would have been drawn to higher $\sigma_8$ by the statistically more powerful $\kappa\kappa$ constraint\footnote{The $\kappa$ auto-spectrum has comparable signal-to-noise ratio to our cross-correlations but obtains better $\sigma_8$ constraints since it does not need to marginalize over bias parameters, e.g.~ref.~\cite{Sailer21}.}.

\begin{figure}[htb]
    \centering
    \resizebox{\columnwidth}{!}{\includegraphics{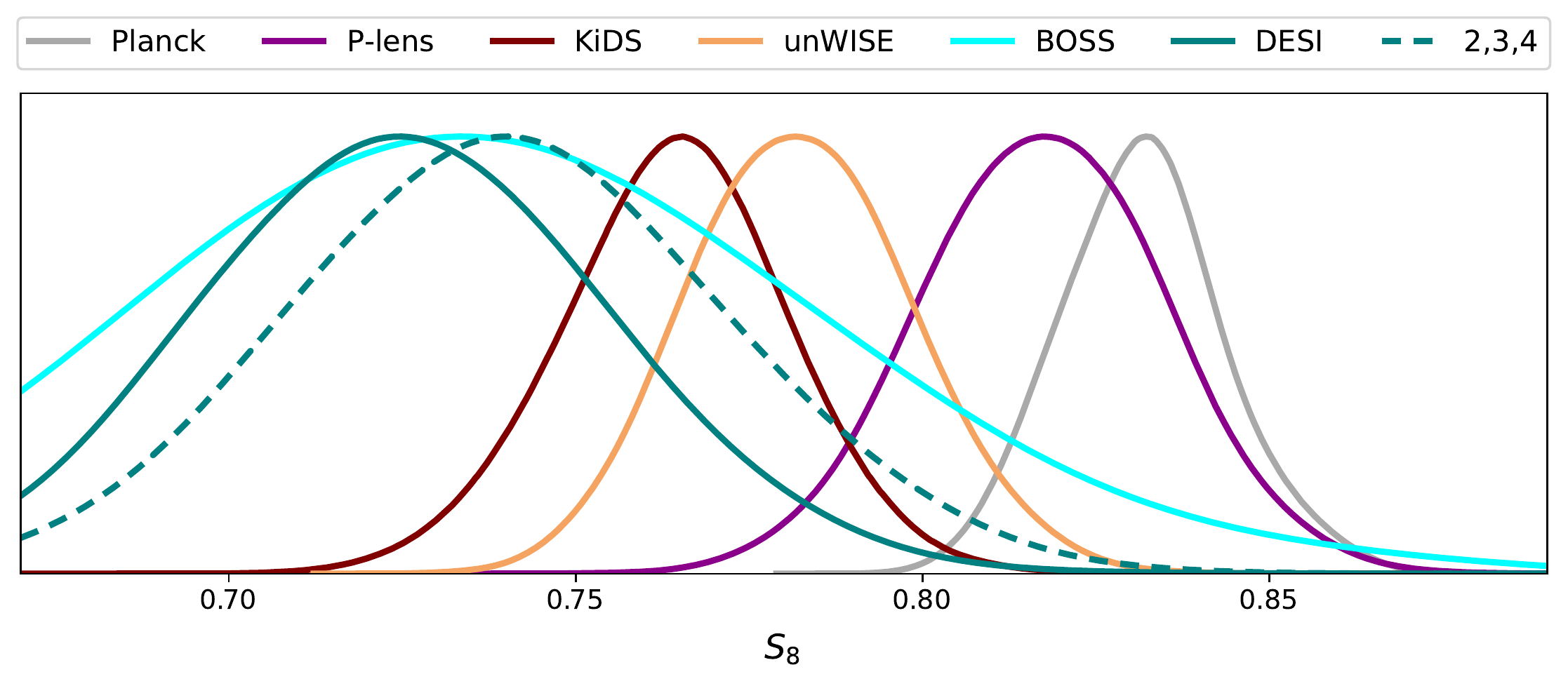}}
    \caption{A comparison of some recent constraints on the clustering amplitude at low redshift.  We show the posterior distribution of $S_8=\sigma_8(\Omega_m/0.3)^{0.5}$, normalized to peak at unity for readability, for a selection of surveys described in the main text.  For the DESI data we show the result using the full sample as a solid line, and omitting the lowest redshift bin as the dashed line. }
    \label{fig:data_cmp_S8}
\end{figure}

Our results reinforce the tendency for (some) lower redshift measurements to have lower matter density or less clustering than inferred by Planck within the context of $\Lambda$CDM.  This is typically summarized in terms of $S_8=\sigma_8(\Omega_m/0.3)^{0.5}$.  In terms of this statistic we find $S_8=0.73\pm 0.03$ for the combined sample, lower than Planck's $S_8=0.832\pm 0.012$.  If we omit the first bin we find $S_8=0.74\pm 0.03$.  Earlier work using similar samples drawn from previous releases of the same imaging data \cite{Hang21a,Hang21b,Kitanidis21} also found lower $S_8$.  Substituting unWISE-selected galaxies for the optical samples also found lower $S_8$ \cite{Krolewski21}.  We are in good agreement with the unWISE analysis ($S_8 = 0.784\pm 0.015$) that used the same CMB lensing maps but a different galaxy sample, extending to higher redshift but with a more uncertain $dN/dz$ and a different theoretical model.  In general we also find good agreement with recent galaxy lensing measurements, which tend to lie lower than the Planck value (Fig.~\ref{fig:data_cmp_S8}).  In particular, we find agreement with the DES Y3 ``$3 \times 2$'' analysis ($S_8 = 0.776\pm 0.017$ \cite{DESY3}), which includes both the auto-correlation of shear\footnote{Adding the Planck lensing auto-correlation to be more consistent with the DES methodology would considerably tighten our contours and move our result upwards, however it would also introduce sensitivity to the higher-redshift matter fluctuations.  We have decided to omit $C_\ell^{\kappa\kappa}$ in order isolate the lower redshifts.} and tracer galaxies, as well as their cross-correlation.  The shear-only analysis \cite{Amon21,Secco21} finds $S_8=0.772\pm0.017$, again in good agreement.  We are also consistent with the lensing and galaxy clustering results from the KiDS+BOSS+2dFlens surveys, with $S_8 = 0.766^{+0.020}_{-0.014}$ \cite{kids1000}, and the reanalysis of KiDS by ref.~\cite{Loureiro21} who find $S_8=0.754\pm 0.028$.  Similarly ``low'' values of $S_8$ have also been found by the first year cosmic shear analysis of HSC \cite{Hikage:2018qbn}, and by CFHTLenS \cite{Heymans:2013fya}.

Hints of a ``low'' value of $\Omega_m$ and $S_8$ in the low-redshift Universe do not rely entirely on lensing.  For example a recent analysis of the ``full-shape'' redshift-space BOSS galaxy power spectrum, BAO and bispectrum \cite{Philcox21} reports $\Omega_m = 0.338^{+0.016}_{-0.017}$ and $\sigma_8 = 0.692^{+0.035}_{-0.041}$, roughly $3.5\sigma$ away from Planck, though this analysis frees the spectral tilt, which is constrained to be $n_s = 0.870^{+0.067}_{-0.064}$; when $n_s$ is fixed to that measured by Planck, the corresponding analysis finds $\Omega_m = 0.320^{+0.010}_{-0.010}$ and $\sigma_8 = 0.722^{+0.032}_{-0.036}$, reducing the tension to less than $3\sigma$. The analysis of ref.~\cite{Philcox21} utilized window-free estimators of galaxy spectra, avoiding a normalization issue that affected some earlier analyses of the BOSS data resulting in artificially low $\sigma_8$ constraints.  Another recent full-shape analysis of BOSS by ref.~\cite{Chen22} finds results lower than but consistent with both the DES Y3 and Planck results (with $\Omega_m = 0.303 \pm 0.008$ and $\sigma_8 = 0.733 \pm 0.047$), using the updated BOSS window functions of ref.~\cite{Beutler21}.  Their fits imply $\Sigma_8=0.74\pm 0.05$, which is in good agreement with our result.  This constraint is shown in Fig.~\ref{fig:cmp_corner}, where we can see that it has a different degeneracy direction than the lensing-based measurements.  Since the model of ref.~\cite{Chen22} is the same as the one employed in our analysis, in future we should be able to replace this BOSS constraint with one from the spectroscopic sample of LRGs produced by DESI for which a joint analysis with our lensing measurement can be performed.  This will allow us to reduce our dependence upon external datasets while retaining tight constraints.  In addition, recent constraints \cite{Zhang21b} from the BOSS correlation function, plus a range of external data, give $\Omega_m=0.306\pm 0.011$ and $\sigma_8=0.766\pm 0.055$; again in good agreement with our results (and also in agreement with Planck).  Finally an analysis of BOSS using an N-body based emulator \cite{Kobayashi21} also returns constraints very close to our result.  By contrast recent analyses of RSD by the eBOSS collaboration find $\sigma_8=0.84\pm 0.06$ \cite{eBOSS:2020yzd} and $\sigma_8=0.814\pm 0.044$ \cite{Semenaite21}, considerably higher than our result though still with large enough error bars that the tension with other results is not high.

In terms of other inferences drawn from galaxies at comparable redshifts, earlier analyses of BOSS galaxies at $z\simeq 0.57$ in combination with Planck lensing \cite{Pullen16,Doux18} have also found lower clustering amplitude than expected from Planck (i.e.\ $C_\ell^{\kappa g}$ is lower on large scales or small $\ell$), though also with large errors so the statistical significance is relatively weak.  Similarly, in an analysis of Planck CMB lensing with BOSS galaxies, ref.~\cite{Singh20} found consistency with Planck, albeit with 10\% lower amplitude. By contrast the earlier Planck analysis \cite{PlanckLens13} and the combination of BOSS galaxies at $z\simeq 0.57$ and the ACT lensing maps, over a much smaller area, finds a consistency with the Planck-preferred normalization \cite{Darwish21}.
In a recent analysis of a wide range of data sets, ref.~\cite{Garcia21} find evidence for a deficit in the growth of large scale structure in the redshift regime $0.2<z<0.6$, which encompasses our lowest redshift bin.  The amplitude of the deficit is smaller than, though consistent with, the shortfall we see.  However much of that shortfall arises from cosmic shear data rather than CMB lensing.  When analyzing their full redshift range and using CMB lensing but not cosmic shear they find agreement with $\Lambda$CDM growth with $S_8=0.825\pm 0.023$, which is higher than our result.  The disagreement is puzzling since the lensing map is the same and much of their constraint comes from a similar redshift range though there are differences in the imaging data used, sample construction and galaxy type, stellar contamination level, $dN/dz$ determination, masks, modeling and nuisance parameters.  It would be interesting to investigate these issues further.

\section{Conclusions}
\label{sec:conclusions}

We have presented constraints on cosmological models from a cross-correlation of Planck CMB lensing \cite{PlanckLens18} and LRGs from the DESI imaging surveys \cite{Dey19}.  We use photometric redshifts to divide the LRGs into 4 samples with mean redshifts of 0.47, 0.63, 0.79 and 0.92.  Our LRG data cover 44\% of the sky ($\approx$18,000 sq.deg.; Fig.~\ref{fig:maps}) and the galaxy auto-spectrum is sample variance limited for all samples on all of the scales that we model.  We report a strong cross-correlation signal in multiple redshift bins (Fig.~\ref{fig:bandpowers}), allowing us to constrain a degenerate combination of the matter density and the amplitude of the matter power spectrum.

Our cross-correlation signal comes primarily from large scales (Fig.~\ref{fig:kernel}) and we model it using a combination of perturbation theory (\S\ref{sec:clpt}) and a Lagrangian bias expansion based on N-body simulations (\S\ref{sec:anzu}).  We show that these models produce unbiased constraints on mock catalogs for the analysis choices we adopt (\S\ref{sec:simulations}; Fig.~\ref{fig:mock_validation}).

Our data alone do not strongly constrain all of the $\Lambda$CDM parameters, so we have held some of them fixed and incorporated other data sets in order to break some degeneracies.  In particular we hold the spectral index ($n_s$), optical depth ($\tau$), baryon density ($\omega_b$) and angular size of the sound horizon ($\theta_\star$) fixed as shown in Table \ref{tab:priors}.  Our results are robust to small changes in these values.  We vary the amplitude of the spectrum ($\ln[10^{10}A_s]$) and the cold dark matter density, plus the bias parameters, counter terms and stochastic terms in the model.  Our data are consistent with a wide range of $\Omega_m$, so to improve convergence and focus attention on the physically interesting region we combine with the Pantheon supernova data and a collection of BAO constraints (but see appendix \ref{app:noext}).  Our results are summarized in Figs~\ref{fig:data_corner} and \ref{fig:sigmaz} and compared to a selection of other measurements in Fig.~\ref{fig:cmp_corner}.  We find results that are lower than the predictions of the $\Lambda$CDM model constrained by the primary CMB anisotropy data as measured by Planck.  Our results also indicate that structure grows with time more slowly than the model predictions, though this is driven primarily by our lowest redshift bin ($z\approx 0.47$) and is of modest significance.

It is of concern that our most discrepant bin is also the one with the lowest signal-to-noise ratio, the smallest dynamic range in scale and the one most sensitive to changes in the Planck lensing map used.  However, despite these factors we have been unable to find a systematic that would account for the lower $\sigma_8$ that we measure.  At present we believe this is most likely to be a statistical fluctuation.  Indeed the best-fitting model with $\Omega_m\simeq 0.31$ and $\sigma_8\simeq 0.81$ (consistent with Planck) provides a statistically acceptable fit to the lowest redshift slice, though the best fit is for lower $\sigma_8$.  Part of the downward shift in $\sigma_8$ for that bin is due to parameter volume effects (mostly $\alpha_\times$) in the marginal posterior due to the limited dynamic range of the data in scale, with the sense that lower $\sigma_8$ correspond to regions where the perturbative model is least reliable (i.e.\ large values of the counterterm).  Further investigation awaits improved modeling.

Our results should have very different systematic uncertainties than other probes.  The galaxy sample and its dependence on observational conditions is well known, and will become increasingly well characterized as the DESI survey proceeds.  The redshift of our source screen (the CMB) is very well known \cite{PlanckLegacy18} and the redshift distribution of our lenses (which are all well separated from the source) has been spectroscopically determined \cite{PaperI}.  As further redshifts are gathered, we will be able to further investigate the dependence of $dN/dz$ on sample selection, imaging properties and photometry.  Spectroscopic redshifts will also allow us to infer the 3D clustering of the imaging sample and, in combination with imaging-spectro cross-correlations, enable us to bound any non-cosmological contribution to $C_\ell^{gg}$ (which could bias our $\sigma_8$ results low).  The availability of accurate photometric redshifts means we can select galaxies to lie in narrow redshift bins, obviating the need to model evolving bias.  We are not affected by intrinsic alignments, blending or other systematics that need to be carefully handled in cosmic shear surveys.  Our signal is dominated by large scales where linear or quasi-linear theory holds and where the impact of baryonic effects is minimal.  As such we provide a valuable check of previous measurements.

Our results are sufficiently tight to begin to constrain the low-redshift amplitude of structure, but fundamentally limited by the CMB lensing maps that we use and by our modeling.  Luckily, we anticipate rapid progress in both directions in the very near future.  Combining more sensitive, ground-based measurements of the CMB with the existing Planck data in maps optimized for cross-correlation should improve the signal to noise on $C_\ell^{\kappa g}$ for all samples.  In combination with models capable of working to higher $\ell$ at low $z$ this will improve the constraints on the power spectrum amplitude, $\sigma_8$.  The well-characterized, wide-area and deep LRG sample that we have used in this analysis, combined with the Lagrangian bias framework, should be ideal for such future work.  Looking forward, DESI will provide spectroscopic redshifts for most of the galaxies used in this work, improving control of systematic errors and allowing us to combine 3D clustering (including RSD and BAO) with lensing for the same sample.  The distance information inherent in the BAO signal will allow us to measure $\Omega_m$ and $h$ internally, while the RSD signal will give us a probe of the amplitude of large-scale structure from the Newtonian potential which can be compared with the Weyl potential measurement from CMB or galaxy-galaxy lensing.

\acknowledgments
We would like to thank David Alonso for help with \texttt{NaMaster}.
We also thank Chris Blake and Alex Krolewski for helpful discussions during the preparation of this manuscript.
M.W.~is supported by the DOE and the NSF.
R.Z.~is supported by the DOE under Contract No. DE-AC02-05CH11231.
J.D.~is supported by the Lawrence Berkeley National Laboratory Chamberlain Fellowship.
S.F.~is supported by the Physics Division of Lawrence Berkeley National Laboratory.
We acknowledge the use of \texttt{NaMaster} \cite{Alonso18}, \texttt{Cobaya} \cite{CobayaSoftware, Torrado21}, \texttt{GetDist} \cite{Lewis19}, \texttt{CAMB} \cite{Lewis00} and \texttt{velocileptors} \cite{Chen20} and thank their authors for making these products public.
This research has made use of NASA's Astrophysics Data System and the arXiv preprint server.
This research is supported by the Director, Office of Science, Office of High Energy Physics of the U.S. Department of Energy under Contract No. DE-AC02-05CH11231, and by the National Energy Research Scientific Computing Center, a DOE Office of Science User Facility under the same contract; additional support for DESI is provided by the U.S. National Science Foundation, Division of Astronomical Sciences under Contract No. AST-0950945 to the NSF's National Optical-Infrared Astronomy Research Laboratory; the Science and Technologies Facilities Council of the United Kingdom; the Gordon and Betty Moore Foundation; the Heising-Simons Foundation; the French Alternative Energies and Atomic Energy Commission (CEA); the National Council of Science and Technology of Mexico; the Ministry of Economy of Spain, and by the DESI Member Institutions.

The authors are honored to be permitted to conduct scientific research on Iolkam Du'ag (Kitt Peak), a mountain with particular significance to the Tohono O'odham Nation.

\bigskip

The data used in this publication are available at \url{https://zenodo.org/record/5834378#.YeBonljMJUQ}

\clearpage

\appendix

\section{Lagrangian perturbation theory formulae}
\label{app:lpt}

For the main results of the paper we have used a variant of Lagrangian perturbation theory, as implemented in the \texttt{velocileptors} code. In the Lagrangian basis galaxy clustering is described by an initial protohalo density, $F(\textbf{q})$, dynamically advected from their initial positions, $\textbf{q}$, to their observed positions, $\textbf{x} = \textbf{q} + \Psi(\textbf{q},t)$, where $\Psi(\textbf{q},t)$ is the displacement, or trajectory, of a galaxy. By number conservation this implies
\begin{equation*}
    1 + \delta_g(\textbf{x},t) = \int d^3\textbf{q}\; F(\textbf{q})\, \delta_D(\textbf{x} - \textbf{q} - \Psi(\textbf{q},t)).
\end{equation*}
The initial density $F(\textbf{q})$ can be written as a bias expansion in the initial conditions; in this paper we adopt the restricted form
\begin{equation*}
    F(\textbf{q}) = 1 + b_1 \delta(\textbf{q}) + \frac{b_2}{2} (\delta(\textbf{q}) - \langle \delta^2 \rangle).
\end{equation*}
The matter overdensity can be expressed in this formalism as an unbiased tracer, i.e.\ $F = 1$. We refer the reader to ref.~\cite{Chen20} for details on higher-order bias contributions since they will not be used in this paper.

The full expression for the galaxy power spectrum in real space is given e.g.\ in \S 4.2.1 of ref.~\cite{Chen20}. For completeness, however, we include in this appendix a table of the relevant contributions to the galaxy autospectrum and galaxy-matter cross spectra in Equation~\ref{eqn:lpt}.
These contributions, for $(i,j) =\{1, b_1, b_2\}$, can each be written as Hankel transforms
\begin{equation}
    P_{ij}(k) = 4 \pi \sum_{n=0}^\infty \int q^2 dq\  e^{-k^2 (X^{\rm lin} + Y^{\rm lin})/2} \left( \frac{kY}{q} \right)^n f^{(n)}_{ij}(q)\ j_n(kq),
\end{equation}
where $j_n$ are spherical Bessel functions and $X$ and $Y$ are the components of the large-scale displacement two-point function (defined below). The functions $f^{(n)}_{ij}(q)$ are given in Table~\ref{tab:fqs} and the angular kernels $K_n$ in these expressions are given by
\begin{equation}
   K^{(n)}_1 =  \frac{q\Theta_{n>0}}{kY^{\rm lin}} \quad , \quad
   K^{(n)}_2 = 1 - \frac{2 n}{k^2 Y^{\rm lin}} \quad , \quad
   K^{(3)}_3 = \left[ 1 - \frac{2 (n-1)}{k^2 Y^{\rm lin}} \right] \frac{q\Theta_{n>0}}{kY^{\rm lin}}.
\end{equation}
where $\Theta_{n>0}$ is zero for $n=0$ and one for $n>0$.
The $X$, $Y$, $U,\ldots$ are correlators of Lagrangian operators. The linear contributions are given by
\begin{align*}
    &X^{\rm lin} = \int \frac{dk}{2\pi^2}\ P_{\rm lin}(k) \Big( \frac{2}{3} - \frac{2}{3}[j_0(kq) + j_2(kq)] \Big) \\
    &Y^{\rm lin} = \int \frac{dk}{2\pi^2}\ P_{\rm lin}(k) \Big( 2j_2(kq) \Big) \\
    &U^{\rm lin} = \int \frac{dk}{2\pi^2}\ k \Big( - P_{\rm lin} j_1(kq) \Big) \\
    &\xi^{\rm lin} = \int \frac{dk}{2\pi^2}\ k^2 P_{\rm lin}(k) j_0(kq).
\end{align*}
The one-loop contributions to the correlators are given by:
\begin{align}
&X^{\rm 1-loop} = 2 X^{(13)} + X^{(22)}, \quad Y^{\rm 1-loop} = 2 Y^{(13)} + Y^{(22)} \nonumber \\
&X^{(22)} = \int \frac{dk}{2\pi^2}\ \frac{9}{98} Q_1(k) \Big( \frac{2}{3} - \frac{2}{3}(j_0(kq) + j_2(kq)) \Big) \nonumber \\
&Y^{(22)} = \int \frac{dk}{2\pi^2}\ \frac{9}{98} Q_1(k)\ 2 j_2(kq) \nonumber \\
&X^{(13)} = \int \frac{dk}{2\pi^2}\ \frac{5}{21} R_1(k) \Big( \frac{2}{3} - \frac{2}{3}(j_0(kq) + j_2(kq)) \Big) \nonumber \\
&Y^{(13)} = \int \frac{dk}{2\pi^2}\ \frac{5}{21} Q_1(k)\ 2 j_2(kq) \nonumber \\
&V = 3 (2 V_1 + V_3)\nonumber \\
&V_1 = \int \frac{dk}{2\pi^2}\ k^{-1}\ \Big(-\frac{3}{7}R_1(k) \Big) j_1(kq) + S(q) \nonumber \\
&V_3 = \int \frac{dk}{2\pi^2}\ k^{-1}\ \Big( -\frac{3}{7}Q_1(k) \Big) j_1(kq)  + S(q) \nonumber \\
&S = \int \frac{dk}{2\pi^2}\ k^{-1}\ \frac{3}{7} \Big( 2R_1 + 4R_2 + Q_1 + 2Q_2 \Big) \frac{j_2(kq)}{kq} \nonumber \\
&T = \int \frac{dk}{2\pi^2}\ k^{-1}\ \left(-\frac{9}{7}\right) \Big( 2R_1 + 4R_2 + Q_1 + 2Q_2 \Big) j_3(kq)  \nonumber \\
&U^{(3)} = \int \frac{dk}{2\pi^2}\ k \Big( -\frac{5}{21}R_1 j_1(kq) \Big) \nonumber \\
&X^{10} = \int \frac{dk}{2\pi^2}\  \frac{1}{7} \Big( 2(R_1-R_2) - (4R_2+2Q_5)j_0(kq) - (3R_1 + 4R_2 + 2Q_5)j_2(kq) \Big)  \nonumber \\
&Y^{10} = \int \frac{dk}{2\pi^2}\  \frac{3}{7} \Big( (3R_1 + 4R_2 + 2Q_5)j_2(kq) \Big) \nonumber \\
&U^{11} = \int \frac{dk}{2\pi^2}\ k \Big( -\frac{6}{7}(R_1+R_2) j_1(kq) \Big) \nonumber \\
&U^{20} = \int \frac{dk}{2\pi^2}\ k \Big( -\frac{3}{7}Q_8 j_1(kq) \Big) \nonumber,
\end{align}
where the functions $R_n$ and $Q_n$ \cite{Matsubara08} are quadratic in (integrals of) $P_{\rm lin}$ whose definitions can be found in Eqs.~(D.3-D.5) in Appendix D of ref.~\cite{Chen20}.

\begin{table}
\begin{center}
\begin{tabular}{c | c }
$(i,j)$ & $f^{(n)}_{ij}(q)$ \\ \hline
$(1,1)$ (Zeldovich) & $1$ \\
$(1,1)$ (1-loop) & $-(k^2/2) (X^{\rm 1-loop} + Y^{\rm 1-loop} K^{(n)}_2) + (k^3/6) (V K^{(n)}_1 + T K^{(n)}_3)$ \\
$(1,b_1)$ & $-2 k (U^{\rm lin} + U^{3}) K^{(n)}_1 - k^2 (X^{10} + Y^{10}K^{(n)}_2)$\\
$(b_1,b_1)$ & $\xi^{\rm lin} - k^2 U_{\rm lin}^2 K^{(n)}_2 - k U^{11} K^{(n)}_1$ \\
$(1,b_2)$ & $-k^2 U_{\rm lin}^2 K^{(n)}_2 - k U^{20} K^{(n)}_1$ \\
$(b_1, b_2)$ &  $- 2 k \xi^{\rm lin} U^{\rm lin} K^{(n)}_1$ \\
$(b_2, b_2)$ & $ \frac{1}{2} \xi_{\rm lin}^2$
\end{tabular}
\caption{Contributions to galaxy-galaxy and galaxy-matter power spectra in LPT.}
\label{tab:fqs}
\end{center}
\end{table}

\section{Constraints without BAO and SNe data}
\label{app:noext}

\begin{figure}
    \centering
    \resizebox{0.7\columnwidth}{!}{\includegraphics{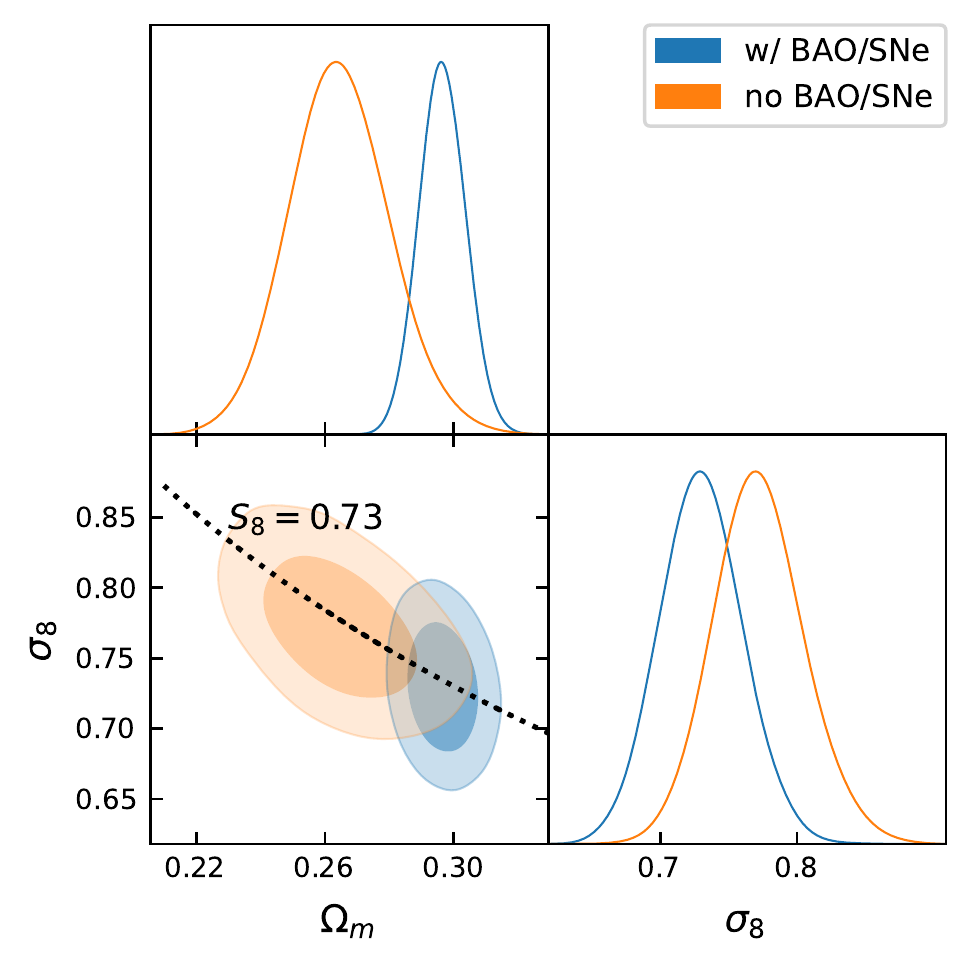}}
    \caption{Marginalized posterior distributions for the matter density, $\Omega_m$, and power spectrum amplitude, $\sigma_8$, for the chains with and without the BAO and SNe data that constrain the distance-redshift relation.  The dotted black line shows $S_8=0.73$. }
    \label{fig:corner_no_ext}
\end{figure}

In the main text (\S\ref{sec:results}) we included constraints on the distance scale from BAO and SNe data (\S\ref{sec:priors}), which together with the angular scale of the CMB serve to tightly constrain $\Omega_m\approx 0.3$.  It is also useful to look at the constraints that come from the lensing and clustering data alone, which we present here.  In order to improve convergence of the chains we tighten the prior on $\alpha_\times$ for the lowest redshift bin sample, which does not have enough dynamic range to constrain $\alpha_\times$ on its own.  Without this tighter prior the chains explore regions where perturbation theory is invalid (i.e.\ the contribution from $\alpha_\times k^2 P(k)$ is a significant contributor to the total power).  Tightening the prior to $\mathcal{N}(0,5)$ for this one parameter keeps the chains in the physical region without leading to bad fits for that sample (modest changes in this prior do not materially change our final constraints since this sample has relatively low statistical weight).  The marginalized posterior distribution for $\Omega_m$ and $\sigma_8$ is shown in Fig.~\ref{fig:corner_no_ext} both with and without the additional data.  The addition of the BAO and SNe data disfavor the low-$\Omega_m$ region that is only weakly constrained by the DESI plus Planck data, but does not appreciably narrow the constraint perpendicular to the $\Omega_m-\sigma_8$ degeneracy.

\clearpage

\bibliographystyle{JHEP}
\bibliography{main}

\end{document}